\documentclass[preprint,showpacs,preprintnumbers,eqsecnum,floatfix]{revtex4}
\usepackage{graphicx}
\usepackage{dcolumn}
\usepackage{bm}

\def\lesssim{\ \raise.3ex\hbox{$<$}\kern-0.8em\lower.7ex\hbox{$\sim$}\ }
\def\gesim{\ \raise.3ex\hbox{$>$}\kern-0.8em\lower.7ex\hbox{$\sim$}\ }
\font\scripti=cmmi7
\font\scriptscripti=cmmi5
\def\sib#1{\setbox0 = \hbox{\scripti #1}
  \kern-.02em\copy0\kern-\wd0
  \kern.04em\box0} 
\def\ssib#1{\setbox0 = \hbox{\scriptscripti #1}
  \kern-.02em\copy0\kern-\wd0
  \kern.04em\box0} 
\font\tenib=cmmib10 
\skewchar\tenib='177 \skewchar\tenib='177 \skewchar\tenib='177
\def\pbold#1{\setbox0 = \hbox{$ #1 $}
  \kern-.022em\copy0\kern-\wd0
  \kern.011em\copy0\kern-\wd0
  \kern.011em\copy0\kern-\wd0
  \kern.011em\copy0\kern-\wd0
  \kern.011em\box0} 
\begin{document}
\title{Single-particle excitations in a trapped
gas of Fermi atoms in the BCS-BEC crossover region}
\author{Y. Ohashi$^{1,2}$ and A. Griffin$^{2}$}
\affiliation{$^1$Institute of Physics, University of Tsukuba, Tsukuba,
  Ibaraki 305, Japan, \\
$^2$ Department of Physics, University of Toronto, Toronto, Ontario, Canada M5S 1A7}
\date{\today}
\begin{abstract}
We investigate the single-particle properties at $T=0$ of a trapped
superfluid gas of Fermi atoms with a Feshbach resonance. 
A tunable pairing interaction associated with the Feshbach resonance 
leads to the BCS-BEC crossover,
where the character of superfluidity
continuously changes from the BCS-type to a BEC of composite bosons
(consisting of a superposition of a condensate of Cooper-pairs and
molecular bosons).
In this paper, we extend our previous work for a uniform superfluid Fermi gas
[Y. Ohashi and A. Griffin, Phys. Rev. A {\bf 67}, 063612 (2003)] to
include the effect of a harmonic trap. 
We do not use the local density approximation (LDA), but directly
solve the Bogoliubov-de Gennes coupled equations. 
Using these equations, we find self-consistent values for the spatially-dependent
local density $n({\bf r})$
as well as the composite BCS order parameter ${\tilde \Delta}({\bf r})$,
the latter describing both the Cooper-pair and molecular condensate
contributions. Using these results, 
we calculate the single-particle density of states in the crossover region, 
and from this 
determine the true single-particle energy gap ($E_{\rm g}$) of the trapped 
Fermi superfluid at $T=0$. This is associated with the in-gap (or Andreev) 
states in the low density region at the edge of the trap. 
We calculate the laser-induced current $I(\omega)$ into another hyperfine
state, as measured in recent rf-spectroscopy experiments. This rf-spectrum
gives a direct probe of the quasiparticle spectrum. 
We show how the high-energy part of $I(\omega)$ gives information about 
${\tilde \Delta}(r=0)$ at the center of the trap (which is comparable to the Fermi
energy $\varepsilon_{\rm F}$ in the crossover region). More generally,
we show that $I(\omega)$ is very dependent on the spatial profile of the pair
potential ${\tilde \Delta}(r)$ which is used. 
We also emphasize that the narrow ``unpaired atom"
peak in the rf-data 
gives information about $E_g$ and the low-energy ($\ll\varepsilon_{\rm F}$)
in-gap states of a Fermi superfluid.
While our calculations are limited at $T=0$, we use them to discuss the recent
Innsbruck data and the LDA calculations of T\"orm\"a and co-workers. The LDA,
while useful, can lead to an incorrect physical picture of the low density surface
region of the Fermi superfluid.
\end{abstract}
\maketitle
%
\section{Introduction}
Recently, the BCS-BEC crossover in a
trapped gas of Fermi atoms near a Feshbach resonance has attracted
much attention\cite{Ohashi1,Ohashi3,Ohashi4,Milstein,Perali}. 
Molecular bosons
associated with a Feshbach resonance can mediate a tunable pairing
interaction between atoms, which becomes stronger with decreasing
threshold energy (denoted by $2\nu$) of the Feshbach
resonance\cite{Timmermans2,Holland1}. The BCS-BEC crossover can be
studied as a function of $2\nu$, with the character of superfluidity
continuously changing from the conventional BCS-type of Cooper-pairs to
a BEC of {\it composite} bosons (consisting of a superposition of 
Cooper-pairs and molecules),
as one approaches the
strong-coupling regime\cite{Ohashi1,Ohashi3,Ohashi4,Milstein}. This situation
contrasts with the ``classic'' crossover physics originally studied in the
superconductivity literature
\cite{Eagles,Leggett,Nozieres,Tokumitsu,Randeria,Melo,Engelbrecht,Haussmann},
where Cooper-pair bosons are always dominant in the whole crossover
regime, from the BCS to the BEC limits. Recent experiments on ultracold gases of 
$^{40}$K and $^6$Li have
used this tunable interaction near the
resonance\cite{Loftus2,Ketterle2}, to produce
a large number of bound states or molecules when $a_s^{2b}>0$ 
($a_s^{2b}$ is the two-body $s$-wave
scattering length)\cite{Regal,Strecker}.
Using this tunable interaction, a BEC of these molecular 
bosons has been observed
in $^{40}$K and $^6$Li\cite{JinA,JochimC}, and more recently,
evidence for superfluidity was found in the BCS side of 
the crossover regime (loosely defined as $a_s^{2b}>0$)
\cite{JinD,Bartenstein,Zwierlein,Kinast,Bourdel}. 
Both single-particle
excitations\cite{Chin} 
and collective 
modes\cite{Kinast,Bar2} have been experimentally studied in the crossover
regime in the case of $^6$Li.
\par
In this paper, we present a detailed study of BCS-type single-particle 
excitations in the BCS-BEC crossover of a trapped gas of Fermi atoms with a
Feshbach resonance. This extends our
previous work for a uniform
gas\cite{Ohashi4} in a major way, since we now include 
the effect of discrete eigenstates due
to confinement in a harmonic trap. 
However, in contrast to Ref.\cite{Ohashi4}, we limit our discussion to $T=0$
in this paper.
Because of the inhomogeneity of a gas
due to the trap potential, the single-particle threshold excitation gap $E_g$ is
no longer simply related to the position-dependent 
superfluid pair potential ${\tilde \Delta}({\bf r})$. This contrasts with
the case of a uniform Fermi superfluid\cite{Ohashi4,Leggett,Randeria}, 
where $E_g$ and the order parameter ${\tilde \Delta}$ are related in a
very direct way ($E_g=|{\tilde \Delta}|$ for $\mu>0$, 
$E_g=\sqrt{\mu^2+|{\tilde \Delta}|^2}$ for $\mu<0$, where $\mu$ is the chemical 
potential of the fermions).  
We compute ${\tilde \Delta}(r)$ self-consistently by solving the
Bogoliubov-de Gennes (BdG) coupled equations and then use it to calculate
the single-particle density of states $N(\omega)$ for a trapped superfluid
gas. We present results through the BCS-BEC crossover by varying the threshold
energy $2\nu$. We also calculate the rf-tunneling current, which can be used to
probe the spectrum of the 
single-particle excitations. We compare our $T=0$ results with 
recent experimental data on $^6$Li\cite{Chin}. 
This data agrees with the calculations of T\"orm\"a
and co-workers\cite{Torma}, using a local density approximation (LDA).
While the LDA is useful, we feel that it gives an incorrect picture of the surface region
of the trapped superfluid.
\par
This paper is organized as follows: In Sec. II, we introduce the usual coupled
fermion-boson Hamiltonian including an isotropic harmonic potential and derive
the mean-filed BdG equations.
The single-particle Green's functions in a trap are 
expressed in terms of the solutions of the BdG equations 
in Sec. III. In Sec. IV, we discuss the BCS-BEC crossover in the cases of 
broad and narrow Feshbach resonances. We review different definitions of the
$s$-wave scattering length. In Sec. V, we calculate the equilibrium properties 
in a self-consistent way, 
such as the chemical potential $\mu$ (which plays a crucial role), the atomic density
profile $n({\bf r})$ and spatial variation of the composite order parameter
${\tilde \Delta}({\bf r})$. The single-particle density of states $N(\omega)$ 
for different values of $2\nu$ 
is calculated and discussed in Sec. VI. In Sec. VII, 
we compare our results (both BdG and LDA) with the 
rf-tunneling spectroscopy data obtained in 
recent experiments on $^6$Li.
Some of our results were briefly reported earlier in Refs.\cite{Ohashi6,Ohashi-web}.
%

%
\par
\vskip5mm
\section{Extension of the Bogoliubov-de Gennes equations ($T=0$)}
\vskip2mm
We consider a two-component Fermi gas with a Feshbach resonance
trapped in a harmonic potential, using the coupled
fermion-boson model
\cite{Ohashi1,Ohashi3,Ohashi4,Milstein,Timmermans2,Holland1,Ranninger}
\begin{eqnarray}
H
&=&
\sum_\sigma\int d{\bf r}
\Psi^\dagger_\sigma({\bf r})
\Bigl[
-{\nabla^2 \over 2m}-\mu+V_{\rm trap}^{\rm F}({\bf r})
\Bigr]
\Psi_\sigma({\bf r})
-U
\int d{\bf r}
\Psi^\dagger_\uparrow({\bf r})
\Psi^\dagger_\downarrow({\bf r})
\Psi_\downarrow({\bf r})
\Psi_\uparrow({\bf r})
\nonumber
\\
&+&
\int d{\bf r}
\Phi^\dagger({\bf r})
\Bigl[
-{\nabla^2 \over 2M}+2\nu-2\mu+V_{\rm trap}^{\rm M}({\bf r})
\Bigr]
\Phi({\bf r})
+
g_{\rm r}
\int d{\bf r}
\Bigl[
\Phi^\dagger({\bf r})\Psi_\downarrow({\bf r})\Psi_\uparrow({\bf r})+h.c.
\Bigr].
\nonumber
\\
\label{eq.2.1}
\end{eqnarray}
Here $\Psi_\sigma({\bf r})$ is the fermion field operator with
pseudo-spin $\sigma=\uparrow,\downarrow$. The Bose quantum field 
operator $\Phi({\bf r})$ 
describes molecular bosons associated with the Feshbach resonance. $U$ is a
non-resonant interaction, which we take attractive ($-U<0$). The Feshbach
resonance $g_{\rm r}$ describes resonance
between one molecule and two Fermi atoms. The effect of this resonance is
controlled by adjusting the energy $2\nu$ of the molecules,
also referred to as the threshold energy. 
(To avoid confusion, we note that in many recent papers, this threshold energy is denoted by $\nu$.)
Since a molecule consists
of two Fermi atoms, we take $M=2m$ and impose the conservation of the
total number of atoms,
\begin{eqnarray}
N
&=&
N_{\rm F}+2N_{\rm M}
\nonumber
\\
&=&
\int d{\bf r} n_{\rm F}(r)
+
2\int d{\bf r} n_{\rm M}(r)
\nonumber
\\
&=&
\sum_\sigma\int d{\bf r}
\langle\Psi^\dagger_\sigma({\bf r})\Psi_\sigma({\bf r})\rangle
+
2\int d{\bf r}\langle\Phi^\dagger({\bf r})\Phi({\bf r})\rangle.
\label{eq.2.2}
\end{eqnarray}
This constraint has already been taken into account in Eq. (\ref{eq.2.1}), with the
Fermi chemical potential $\mu$ and Bose chemical potential
$\mu_M\equiv2\mu$\cite{Ohashi1}. $V_{\rm trap}^{\rm F}({\bf r})$ and
$V_{\rm trap}^{\rm M}({\bf r})$ are the harmonic trap potentials for Fermi
atoms and Bose molecules, respectively, which are assumed to be isotropic 
\begin{equation}
V_{\rm trap}^{\rm F}={1 \over 2}m\omega_0^2r^2,~~~
V_{\rm trap}^{\rm M}={1 \over 2}M\omega_{0M}^2r^2.
\label{eq.2.3}
\end{equation}
In addition, in this paper, we assume that the atoms and molecules feel 
the same trap frequency $\omega_0=\omega_{0M}$ (which correctly describes 
recent experiments).
\par
To study the BCS-BEC crossover phenomenon, we extend
the theory at $T=0$ developed 
by Leggett\cite{Leggett} 
in the context of superconductivity
to include a Feshbach resonance as well as the effect 
of a harmonic trap potential. 
The key point of this theory is 
to solve the mean-field gap equation for the order
parameter together with the equation for the number 
of particles, which determines
the Fermi chemical potential $\mu$. In the BCS-BEC crossover, $\mu$
decreases from the usual BCS limit given by the Fermi energy 
$\varepsilon_{\rm F}$ and can become negative. The {\it thermal} 
fluctuations in the Cooper-channel and the {\it thermal}
excitations of Bose condensate fluctuations, crucial in considering the
BEC-BEC crossover 
at finite temperatures close to 
$T_{\rm c}$\cite{Nozieres,Ohashi1,Ohashi3,Ohashi4,Milstein,Randeria,Melo,Engelbrecht,Haussmann}, are not important at $T=0$. Such fluctuations will not be considered in the present paper.
\par
The gap equation is obtained from the
mean-field approximation for Eq. (\ref{eq.2.1})
in terms of the BCS Cooper-pair condensate
$\Delta({\bf r})\equiv U\langle\Psi_\downarrow({\bf r})
\Psi_\uparrow({\bf r})\rangle$ as well as the molecular BEC 
condensate $\phi_M({\bf r})\equiv\langle\Phi({\bf r})\rangle$. The 
Hartree-Fock-Bogoliubov (HFB) mean-field Hamiltonian for the 
{\it fermions} is given by
\begin{eqnarray}
H_{\rm HFB}
&=&
\sum_\sigma
\int d{\bf r}
{\hat \Psi}_\sigma^\dagger({\bf r})
\Bigl[
-{\nabla^2 \over 2m}-{U \over 2}n_{\rm F}
({\bf r})+V_{\rm trap}^{\rm F}(r)-\mu
\Bigr]
\Psi_\sigma({\bf r})
\nonumber
\\
&-&
\int d{\bf r}\Delta({\bf r})
\Bigr[
{\hat \Psi}_\uparrow^\dagger({\bf r})
{\hat \Psi}_\downarrow^\dagger({\bf r})
+h.c.
\Bigr]
+g_{\rm r}
\int d{\bf r}\phi_M({\bf r})
\Bigr[
{\hat \Psi}_\uparrow^\dagger({\bf r})
{\hat \Psi}_\downarrow^\dagger({\bf r})
+h.c.
\Bigr]
\nonumber
\\
&=&
\sum_\sigma
\int d{\bf r}
{\hat \Psi}_\sigma^\dagger({\bf r})
\Bigl[
-{\nabla^2 \over 2m}
-{U \over 2}n_{\rm F}({\bf r})+V_{\rm trap}^{\rm F}(r)-\mu
\Bigr]
\Psi_\sigma({\bf r})
\nonumber
\\
&-&
\int d{\bf r}{\tilde \Delta}({\bf r})
\Bigr[
{\hat \Psi}_\uparrow^\dagger({\bf r})
{\hat \Psi}_\downarrow^\dagger({\bf r})
+h.c.
\Bigr].
\label{eq.2.6}
\end{eqnarray}
In Eq. (\ref{eq.2.6}), the Cooper-pair order parameter $\Delta({\bf
  r})$ and the 
molecular condensate $\phi_M({\bf r})$ are taken to be real, without loss of 
generality. $n_{\rm F}({\bf r})\equiv\sum_\sigma
\langle\Psi^\dagger_\sigma({\bf r})\Psi_\sigma({\bf r}))\rangle$ 
is the local number density 
of Fermi atoms. 
Equation (\ref{eq.2.6}) clearly has the same form as the
usual BCS-Gor'kov Hamiltonian, but now with the Cooper-pair order parameter 
$\Delta({\bf r})$ replaced with the {\it composite} order parameter,
\begin{equation}
{\tilde \Delta}({\bf r})\equiv\Delta({\bf r})-g_{\rm r}\phi_M({\bf
  r}).
\label{eq.2.7}
\end{equation}
Because of the spherical symmetry of our harmonic trap, 
$\Delta({\bf r})$, $\phi_M({\bf r})$, ${\tilde \Delta}({\bf r})$, 
and $n_{\rm F}({\bf r})$ only depend on $|{\bf r}|$.
\par
Since we are only discussing $T=0$ in this paper, the molecules described
by $\Phi({\bf r})$ are all Bose-condensed. As noted above, we
ignore excitations of this molecular condensate.
Thus the HFB mean-field Hamiltonian in Eq. (\ref{eq.2.6}) 
does not involve the dynamics of the
molecular condensate.
The latter only enters through the equilibrium value ${\tilde \Delta}({\bf r})$.
\par
Superfluidity in our coupled fermion-boson model is characterized by
the two broken-symmetry order parameters, i.e., the BCS order 
parameter 
$\Delta({\bf r})=U\langle\Psi_\downarrow({\bf r})
\Psi_\uparrow({\bf r})\rangle$ 
and the molecular BEC 
condensate
$\phi_M({\bf r})=\langle\Phi({\bf r})\rangle$. However, 
these are strongly
coupled to each other and hence are not independent. 
One finds from Eq. (\ref{eq.2.1}) that
their equilibrium values satisfy
\begin{equation}
{g_{\rm r} \over U}\Delta({\bf r})+
\Bigl[
-{\nabla^2 \over 2M}+V_{\rm trap}^{\rm M}(r)+2\nu-2\mu
\Bigr]\phi_M({\bf r})=0.
\label{eq.2.4}
\end{equation} 
This is in fact an exact identity, and can be obtained from the
equation of motion
\begin{equation}
0={d\phi_M({\bf r},t) \over dt}
={i \over \hbar}
\langle
[H,\Phi({\bf r},t)]
\rangle.
\label{eq.2.5}
\end{equation}
In the absence of a trap, Eq. (\ref{eq.2.4}) reduces to
\begin{equation}
{g_{\rm r} \over U}\Delta+(2\nu-2\mu)\phi_M=0,
\label{eq.2.5b}
\end{equation}
a result discussed at length in Refs. \cite{Ohashi1,Ohashi4}.
As a result of this strong coupling, 
the Cooper-pair and Feshbach molecule condensates are hybridized
by the Feshbach resonance, and are both finite throughout the superfluid phase. 
\par
In a uniform gas, 
the composite order parameter appearing in Eq. (\ref{eq.2.6}) and defined 
in Eq. (\ref{eq.2.7}) can be written in the form
\begin{equation}
{\tilde \Delta}=U_{\rm eff}\sum_{\bf p}
\langle
c_{-{\bf p}\downarrow}c_{{\bf p}\uparrow}
\rangle,
\label{eq.2.5c}
\end{equation}
where $c_{{\bf p}\sigma}$ 
the annihilation operator of a Fermi atom in momentum space,
and\cite{Ohashi1,Ohashi4} 
\begin{equation}
U_{\rm eff}\equiv U+{g_{\rm r}^2 \over 2\nu-2\mu}.
\label{eq.2.5d}
\end{equation} 
$U_{\rm eff}$ describes an effective pairing interaction in a BCS-type
Hamiltonian in Eq. (\ref{eq.2.6}), which can be tuned by
adjusting the molecular threshold energy $2\nu$.
We note that the expression in Eq. (\ref{eq.2.5c}) has the same form
as the usual definition of a
BCS Cooper-pair, apart from the effective pairing interaction $U_{\rm eff}$.
\par
The molecular Bose excitations are described by the field operator
$\delta\Phi({\bf r})\equiv\Phi({\bf r})-\langle\Phi({\bf
  r})\rangle=\Phi({\bf r})-\phi_m({\bf r})$. When we substitute this
expression into the boson kinetic term [the bilinear term involving $\Phi({\bf r})$] 
in Eq. (\ref{eq.2.1}), the terms linear in
 $\delta\Phi({\bf r})$ are cancelled out by another
term which is linear in $\delta\Phi({\bf r})$ arising from the last term in
Eq. (\ref{eq.2.1}). This is a consequence of the key relation in
Eq. (\ref{eq.2.4}). This leaves a term in Eq. (\ref{eq.2.1}) which is bilinear in
$\delta\Phi({\bf r})$, namely
\begin{equation}
H_{\rm M}=
\int d{\bf r}
\delta\Phi^\dagger({\bf r})
\Bigl[
-{\nabla^2 \over 2M}+2\nu+V_{\rm trap}^{\rm M}(r)-2\mu
\Bigr]\delta\Phi({\bf r}).
\label{eq.2.6b}
\end{equation}
Equation (\ref{eq.2.6b}) gives the lowest molecular excitation energy 
$E^{\rm M}_0\equiv 2\nu+(3/2)\omega_{0}-2\mu$. 
A self-consistent calculation shows
that this threshold energy
is always {\it positive} (i.e., $2\nu+(3/2)\hbar\omega_0>2\mu$). 
As discussed in Ref. \cite{Ohashi4}
for the uniform case, this contradicts the fact that excitation
spectrum must be {\it gapless} in a uniform interacting Bose gas (of molecules). 
Although
Eq. (\ref{eq.2.1}) does not explicitly involve an interaction between the
molecules, an effective repulsive interaction 
is induced by the Fermi gas\cite{Ohashi4,Strinati,Petrov}. 
This effective repulsive interaction will lead
to a Bogoliubov phonon as the collective mode in the molecular gas in the BEC regime.
\par 
In an interacting Bose gas, it is well known that the only low-energy 
excitations are collective modes. 
This is because the single-particle
excitations are strongly hybridized\cite{Griffin} with the two-particle 
excitations (including collective modes). 
When we include
the effective molecule-molecule 
interaction in a consistent way,
the gapless or phonon behavior of molecular Bose excitations in a uniform gas
must be recovered. 
This requires an extended version of the present theory, as 
we shall discuss in a future paper\cite{Ohashi7}.
The present paper is mainly concerned with the single-particle 
excitations of a Fermi superfluid, and does not deal with the collective modes.
The fact that our present treatment does not include the correct 
interaction\cite{Strinati,Petrov}
between molecules in the BEC limit is of little importance when discussing 
the spectrum of the single-particle Fermi excitations
(which disappear as we approach the BEC limit).
\par
The mean-field HFB Hamiltonian in
Eq. (\ref{eq.2.6}) is formally identical to a trapped
superfluid Fermi gas in the standard BCS treatment, apart from the
replacement of $\Delta({\bf r})$ by the (self-consistent) composite order parameter
${\tilde \Delta}({\bf r})$. Bruun and co-workers have presented
detailed numerical results for the Cooper pair order
parameter and the single-particle BCS excitations in a BCS superfluid
at $T=0$\cite{Bruun0,BruunA,BruunB,BruunC}. Our present work may be
viewed as a natural
extension to include the effect of the Feshbach resonance, based on
the model in Eq. (\ref{eq.2.6}) involving the effective pair potential 
${\tilde \Delta}({\bf
  r})$. The latter, of course, involves the molecular condensate
$\phi_M({\bf r})$ and must be
computed self-consistently.
\par
As usual, the HFB Hamiltonian in Eq. (\ref{eq.2.6}) can be diagonalized as
$H_{\rm HFB}=\sum_{n\sigma}E_n\gamma_{n\sigma}^\dagger\gamma_{n\sigma}$
by solving the 
Bogoliubov de Gennes (BdG) equations\cite{BdG},
\begin{eqnarray}
\left(
\begin{array}{cc}
H_0&{\tilde \Delta}({\bf r})\\
{\tilde \Delta}({\bf r})&-H_0
\end{array}
\right)
\left(
\begin{array}{c}
u_n({\bf r})\\
v_n({\bf r})
\end{array}
\right)
=
E_n
\left(
 \begin{array}{c}
u_n({\bf r})\\
v_n({\bf r})
\end{array}
\right),
\label{eq.2.5e}
\end{eqnarray}
where $H_0$ is the diagonal component of $H_{\rm HFB}$.
The Bogoliubov quasiparticle excitations are described 
by the fermion operators
$\gamma_{n\sigma}$, which are related to the fermion 
field operator $\Psi_\sigma({\bf r})$ as\cite{BdG}
\begin{eqnarray}
\begin{array}{l}
\displaystyle
\Psi_\uparrow({\bf r})=\sum_n
\Bigl[
u_n({\bf r})\gamma_{n\uparrow}+
v_n^*({\bf r})\gamma^\dagger_{n\downarrow}
\Bigr],\\
\displaystyle
\Psi_\downarrow({\bf r})=\sum_n
\Bigl[
u_n({\bf r})\gamma_{n\downarrow}-
v_n^*({\bf r})\gamma^\dagger_{n\uparrow}
\Bigr].\\
\end{array}
\label{eq.2.5f}
\end{eqnarray}
\par
As noted above, the solutions of these 
BdG equations for a trapped Fermi gas have been discussed
extensively by Bruun and co-workers\cite{Bruun0,BruunA,BruunB,BruunC}
in the BCS limit and where ${\tilde \Delta}({\bf r})=\Delta({\bf r})$. It is
convenient to expand the fermion field operator $\Psi_\sigma({\bf r})$
with respect to 
the eigenfunctions of a harmonic potential 
$V_{\rm trap}^F(r)$ as
\begin{equation}
{\hat \Psi}_\sigma({\bf r})=\sum_{nlm}f^F_{nlm}({\bf r})c_{nlm\sigma}.
\label{eq.2.10b}
\end{equation}
Here, $f^F_{nlm}({\bf r})\equiv R^F_{nl}(r)Y_{lm}({\hat \theta})$, 
where $Y_{lm}({\hat \theta})$ is a spherical harmonic and
$R^F_{nl}(r)$ is the usual radial wavefunction, given by
\begin{equation}
R^F_{nl}(r)=
\sqrt{2}(m\omega_0)^{3/4}
\sqrt{n! \over (n+l+1/2)!}
e^{-{{\bar r}^2 \over 2}}
{\bar r}^l
L_n^{l+1/2}({\bar r}^2)~~~({\bar r}\equiv\sqrt{m\omega_0}r),
\label{eq.2.11}
\end{equation}
where $L_n^{l+1/2}({\bar r}^2)$ is a Laguerre polynomial. The
HFB Hamiltonian $H_{\rm HFB}$ in
Eq. (\ref{eq.2.6}) can then be reduced to
\begin{eqnarray}
H_{\rm HFB}
=\sum_{nlm,\sigma}\xi_{nl}^{\rm F}c^\dagger_{nlm\sigma}c_{nlm\sigma}
&-&
{U \over 2}
\sum_{nn'lm,\sigma}J_{nn'}^l
c^\dagger_{nlm\sigma}c_{n'lm\sigma}
\nonumber
\\
&-&
\sum_{nn'lm}F_{nn'}^l
\Bigl[
c^\dagger_{nlm\uparrow}c^\dagger_{n'l,-m\downarrow}+h.c.
\Bigr].
\label{eq.2.12}
\end{eqnarray}
Here $\xi^{\rm F}_{nl}=\hbar\omega_0(2n+l+3/2)-\mu$ 
are the single-particle excitation energies of the atoms in
the harmonic potential. $F_{nn'}^l$ and $J_{nn'}^l$ 
describe the mean field effects associated with the composite pair potential
 ${\tilde \Delta}(r)$ (which plays the role of an ``off-diagonal" potential) 
and the Hartree potential 
$-{U \over 2}n_{\rm F}(r)$, respectively. 
These are given by
\begin{equation}
F_{nn'}^l\equiv\int_0^\infty
 r^2dr R^F_{nl}(r){\tilde \Delta}(r)R^F_{n'l}(r),
\label{eq.2.13}
\end{equation}
\begin{equation}
J_{nn'}^l\equiv\int_0^\infty
 r^2dr R^F_{nl}(r)n_{\rm F}(r)R^F_{n'l}(r).
\label{eq.2.14}
\end{equation}
We note that $F_{nn'}^l$ and $J_{nn'}^l$ include both the
intra-shell terms ($n=n'$) as well as the inter-shell terms ($n\ne n'$). The
local Cooper-pair order parameter and fermion density are given by\cite{Bruun0}
\begin{eqnarray}
\Delta(r)
=
U
\sum_{nn'l}
{2l+1 \over 4\pi}
R^F_{nl}(r)R^F_{n'l}(r)
\langle c_{nl0\downarrow}c_{n'l0\uparrow}\rangle,
\label{eq.2.15}
\end{eqnarray}
\begin{eqnarray}
n_F(r)
=
\sum_{nn'l\sigma}
{2l+1 \over 4\pi}
R^F_{nl}(r)R^F_{n'l}(r)\langle
c^\dagger_{nl0\sigma}c_{n'l0\sigma}\rangle.
\label{eq.2.16}
\end{eqnarray}
In Eq. (\ref{eq.2.16}), we have taken advantage of the spherical symmetry of our
model, which leads to
$\langle c_{nlm\downarrow}c_{n'l,-m\uparrow}\rangle
=\langle c_{nl0\downarrow}c_{n'l0\uparrow}\rangle$ and
$\langle c^\dagger_{nlm\sigma}c_{n'lm\sigma}\rangle
=\langle c^\dagger_{nl0\sigma}c_{n'l0\sigma}\rangle$. 
\par
It is important to remember that Eq. (\ref{eq.2.12}) 
includes the effect of the molecular condensate
$\phi_M(r)$, since it enters the composite order parameter ${\tilde
  \Delta}(r)=\Delta(r)-g_{\rm r}\phi_M(r)$. Since
$\phi_M(r)$ only depends on $r$, we need only consider the $l=0$
quantum number. Thus we can expand $\phi_M(r)$ in terms of the radial components
\begin{eqnarray}
\phi_M(r)={1 \over \sqrt{4\pi}}\sum_n\alpha_n R_{n0}^M(r).
\label{eq.2.17}
\end{eqnarray}
Here $f^M_{nlm}({\bf r})\equiv R^M_{nl}(r)Y_{lm}({\hat \theta})$ is a
molecular eigenfunction for 
the isotropic harmonic potential $V_{\rm trap}^{\rm M}(r)$,
with the molecular energy $E^M_{nl}=\hbar\omega_{0}(2n+l+3/2)$. The radial
component $R^M_{nl}(r)$ is identical to that given in Eq. (\ref{eq.2.11}),
except that the atom mass $m$ is now replaced by the bound state mass $M=2m$.
Substituting Eq. (\ref{eq.2.17}) into Eq. (\ref{eq.2.4}), we obtain an 
expression for $\alpha_n$ in terms of $\Delta(r)$,
\begin{eqnarray}
\alpha_n=
-{g_{\rm r} \over U}
{\sqrt{4\pi} \over E_{n0}^M+2\nu-2\mu}
\int_0^\infty dr r^2\Delta(r) R_{n0}^M(r).
\label{eq.2.18}
\end{eqnarray}
The magnitude of the various expansion coefficients $\alpha_n$ clearly determines to
what extent the molecular
condensate $\phi_M(r)$ in Eq. (\ref{eq.2.17}) is similar to the BEC order parameter 
$\phi_M^{\rm ideal}(r)$ of a non-interacting Bose gas of molecules 
in a harmonic
trap. The latter is given by the macroscopic occupation of the lowest ($n=0$) state
\begin{equation}
\phi_M^{\rm ideal}(r)={1 \over \sqrt{4\pi}}\alpha_0 R_{00}^M(r),
\label{eq.2.18b}
\end{equation}
where $|\alpha_0|=\sqrt{N/2}$ for a non-interacting Bose gas. 
\par
Since Eq. (\ref{eq.2.12}) can be written as $H_{\rm
  HFB}=\sum_{ml}H_{\rm HFB}^{ml}$, one may independently diagonalize
each $H_{\rm HFB}^{ml}$, i.e., for each set of values $(l,m)$, by a
unitary transformation. This is the Bogoliubov 
transformation,
\begin{eqnarray}
\left(
\begin{array}{c}
c_{0lm\uparrow} \\
\ldots \\
c_{N_llm\uparrow} \\
c^\dagger_{0l,-m\downarrow} \\
\ldots \\
c^\dagger_{N_ll,-m\downarrow} \\
\end{array}
\right)=
{\hat W}^l
\left(
\begin{array}{c}
\gamma_{0lm\uparrow} \\
\ldots \\
\gamma_{N_llm\uparrow} \\
\gamma^\dagger_{0lm\downarrow} \\
\ldots \\
\gamma^\dagger_{N_llm\downarrow} \\
\end{array}
\right).
\label{eq.2.22}
\end{eqnarray}
Here ${\hat W}^l$ is a $2(N_l+1)\times 2(N_l+1)$-orthogonal
matrix, and the fermion operator $\gamma_{nlm\sigma}$ describes the Bogoliubov 
quasi-particles. Physically, the matrix elements of ${\hat W}^l$
describe the hybridization of particle and hole excitations in
the Bogoliubov quasi-particles described by
$\gamma_{nlm\sigma}$. ${\hat W}^l$ is determined by the requirement that
${\hat W}^{l\dagger} H_{\rm HFB}^{ml} {\hat W}^l$ be diagonal.
The matrix elements $W^l_{ij}$ are then obtained
from the eigenfunctions of the following BdG equations for $H_{\rm
  HFB}^{ml}$\cite{note4},
\begin{eqnarray}
\left(
\begin{array}{cccccc}
\xi^F_{0l}-{U \over 2}J_{00}^l & \ldots & -{U \over 2}J_{0N_l}^l &
-F_{00}^l & \ldots & -F_{0N_l}^l \\
\ldots & \ldots & \ldots & \ldots & \ldots & \ldots \\
-{U \over 2}J_{N_l0}^l & \ldots & \xi^F_{N_ll}-{U \over 2}J_{N_lN_l}^l & 
-F_{N_l0}^l & \ldots & -F_{N_lN_l}^l \\
-F_{00}^l & \ldots & -F_{0N_l}^l &
-\xi^F_{0l}+{U \over 2}J_{00}^l & \ldots & {U \over 2}J_{0N_l}^l \\
\ldots & \ldots & \ldots & \ldots & \ldots & \ldots \\
-F_{N_l0}^l & \ldots & -F_{N_lN_l}^l &
{U \over 2}J_{N_l0}^l & \ldots & -\xi^F_{N_ll}+{U \over 2}J_{N_lN_l}^l
\end{array}
\right)
\left(
\begin{array}{c}
W^l_{0,n} \\
\ldots \\
W^l_{N_l,n} \\
W^l_{N_l+1,n} \\
\ldots \\
W^l_{2N_l+2,n} \\
\end{array}
\right)
{~}
\nonumber
\\
&{~}&
\hskip-40mm
=E^F_{nl}
\left(
\begin{array}{c}
W^l_{0,n} \\
\ldots \\
W^l_{N_l,n} \\
W^l_{N_l+1,n} \\
\ldots \\
W^l_{2N_l+2,n} \\
\end{array}
\right).
\label{eq.2.21}
\end{eqnarray}
As usual, we introduce a cutoff $\omega_c\equiv \hbar\omega_0(N_c+3/2)$ in
the energy summation in the gap equation [see Eq. (\ref{eq.2.23}) below]. 
This defines $N_c$.
The maximal radial quantum number $N_l$ in
Eq. (\ref{eq.2.21}) is then given by the largest integer bounded by $(N_c-l)/2$. 
As with the usual BdG equations, positive and
negative eigenenergies ($E_{nl}^F$ and $-E_{nl}^F$) are obtained from
Eq. (\ref{eq.2.21}). The diagonalized Hamiltonian can be written
using only the positive energy eigenenergies ($E^F_{nl}\ge 0$), namely
\begin{eqnarray}
H_{\rm HFB}^F
&=&
\sum_{n=0}^{N_l}
\Bigl[
E_{nl}^F\gamma_{nlm\uparrow}^\dagger\gamma_{nlm\uparrow}-
E_{nl}^F\gamma_{nlm\downarrow}\gamma^\dagger_{nlm\downarrow}
\Bigr]
\nonumber
\\
&=&
\sum_{n=0}^{N_l}E_{nl}^F+
\sum_{n=0,\sigma}^{N_l}E^F_{nl}\gamma_{nlm\sigma}^\dagger\gamma_{nlm\sigma}.
\label{eq.2.22b}
\end{eqnarray}
\par
In the following discussion, we only take the {\it positive} eigenenergies
$E^F_{nl}$ as in Eq. (\ref{eq.2.22b}).
According to Eq. (\ref{eq.2.22b}), 
the operator $\gamma^\dagger_{nlm\sigma}$ describes
creating a Fermi single-particle 
excitation from the ground state, with excitation energy 
$E_{nl}^F\ge 0$. 
Because of the assumed spherical symmetry of our trap, 
$E^F_{nl}$ and ${\hat W}^l$ do not depend on the quantum
number $m$. Substituting Eq. (\ref{eq.2.22}) into Eqs. (\ref{eq.2.15})
and (\ref{eq.2.16}), we obtain (at $T=0$)
\begin{eqnarray}
\Delta(r)=U
{\sum_{nn'l}}'{2l+1 \over 4\pi}
R^F_{nl}(r)R^F_{n'l}(r)
\sum_{j=0}^{N_l}
W_{{\bar N}_l+n,{\bar N}_l+j}^l
W_{n'+1,{\bar N}_l+j}^l,
\label{eq.2.23}
\end{eqnarray}
\begin{eqnarray}
n_{\rm F}(r)=
\sum_{nn'l}{2l+1 \over 4\pi}
R^F_{nl}(r)R^F_{n'l}(r)
\sum_{j=0}^{N_l}
\Bigl[
W_{n+1,{\bar N}_l+j}^l
W_{n'+1,{\bar N}_l+j}^l
+
W_{{\bar N}_l+n,j+1}^l
W_{{\bar N}_l+n',j+1}^l
\Bigr],
\label{eq.2.24}
\end{eqnarray}
where ${\bar N}_l\equiv N_l+2$ and the prime in  ${\sum}'$ refers to
the finite cutoff $\omega_c$ in the summation (see above). 
\par
In the BCS-BEC crossover region, we shall find that 
the Fermi chemical potential $\mu$
deviates strongly from the usual BCS limit, where it equals the Fermi energy 
$\varepsilon_{\rm F}$. This effect is taken into account by
considering the
equation for the number of atoms in addition to the 
BdG equations\cite{Leggett}.
Since non-condensed molecules are absent at $T=0$,
the {\it total} number density of Fermi atoms $n(r)$ is simply given by
\begin{equation}
n(r)=2|\phi_M(r)|^2+n_{\rm F}(r)\equiv 2n_{\rm M}(r)+n_{\rm F}(r),
\label{eq.2.25}
\end{equation}
where each molecule counts for two Fermi atoms. The equation for the
total number of atoms, which determines $\mu$, is then obtained by
integrating Eq. (\ref{eq.2.25}) over ${\bf r}$. The result is
\begin{eqnarray}
N
&=&
2\sum_n\alpha_n^2
+\sum_{nn'l}(2l+1)
\Bigl[
|W_{n+1,{\bar N}_l+n'}^l|^2+
|W_{{\bar N}_l+n,n'+1}^l|^2
\Bigr]
\nonumber
\\
&\equiv&
2N_{\rm M}+N_{\rm F}.
\label{eq.2.26}
\end{eqnarray} 
Here $N_{\rm M}$ and $N_{\rm F}$ are the number of (Feshbach) molecular bosons and number of Fermi atoms in the crossover region, respectively.
Equations (\ref{eq.2.21}) and Eq. (\ref{eq.2.26}) are the basic equations 
of our theory at $T=0$, taking into account
a Bose condensate of Cooper-pairs and molecules 
in a self-consistent manner. We numerically solve
the BdG equations (\ref{eq.2.21}) together with the generalized number
equation in Eq. (\ref{eq.2.26}), determining the coefficient
$\alpha_n$ [see Eq. (\ref{eq.2.18})], $\Delta(r)$, $\phi_M(r)$,
$n_F(r)$ and $\mu$ self-consistently. 
\par
In order to understand the essential physics, we end this Section by recalling 
what the above formalism reduces to for a {\it uniform} superfluid 
Fermi gas. In
this case, the simple pairing mean-field approximation (MFA) gives
usual BCS-Gor'kov expressions for 
the diagonal and off-diagonal single-particle Green's functions
\begin{eqnarray}
G_{11}({\bf p},\omega)
=
{\omega+\xi_{\bf p} \over \omega^2-E_{\bf p}^2},
~~~G_{12}({\bf p},\omega)
=
-{{\tilde \Delta} \over \omega^2-E_{\bf p}^2},
\label{eq.10}
\end{eqnarray}
with the BCS-Bogoliubov excitation energy given by
\begin{equation}
E_{\bf p}=\sqrt{(\varepsilon_{\bf p}-\mu)^2+|{\tilde \Delta}|^2}.
\label{eq.11}
\end{equation}
Here $\xi_{\bf p}\equiv\varepsilon_{\bf p}-\mu$ (where 
$\varepsilon_{\bf p}=p^2/2m$) is the kinetic energy 
measured from the chemical potential.
Using $G_{12}({\bf p},\omega)$ to calculate $\Delta$, 
one finds that ${\tilde \Delta}$ in Eq. (\ref{eq.2.5c})
satisfies the usual ``gap equation" [but now with the pairing interaction $U_{\rm
  eff}$ given by Eq. (\ref{eq.2.5d})],
\begin{equation}
{\tilde \Delta}=U_{\rm eff}
{\sum_{\bf p}}'
{{\tilde \Delta} \over 2E_{\bf p}}\tanh{1 \over 2}\beta E_{\bf p},
\label{eq.13}
\end{equation}
valid at finite temperatures.
As usual, a cutoff is introduced in the momentum
summation. At 
this MFA level, the total number of atoms at $T=0$ has 
the same form as Eq. (\ref{eq.2.26}),
\begin{equation}
N=2N_{\rm M}+N_{\rm F}.
\label{eq.13b}
\end{equation}
The number of Fermi atoms in Eq. (\ref{eq.13b}) 
is now given [using $G_{11}({\bf p},\omega)$]
by the well-known BCS expression
\begin{equation}
N_{\rm F}=
\sum_{{\bf p},\sigma}
\langle c_{{\bf p}\sigma}^\dagger c_{{\bf p}\sigma}
\rangle
=
\sum_{\bf p}
\Bigl[
1-{\xi_{\bf p} \over E_{\bf p}}\tanh{1 \over 2}\beta E_{\bf p}
\Bigr],
\label{eq.12}
\end{equation}
while the number of condensed molecules is $N_{M}=|\phi_M|^2$.
\par
In the uniform gas, the
 values of $\mu$ and ${\tilde \Delta}$ are determined by the
self-consistent solutions of the MFA gap equation ({\ref{eq.13}) and
  the number equation given by Eqs. (\ref{eq.13b}).
This simple ``pairing approximation" for the single-particle Fermi excitations 
is expected to give a quantitative description at $T=0$ [where $\tanh
{1 \over 2}\beta E_{\bf p}\to 1$ in Eqs. (\ref{eq.13}) and (\ref{eq.12})], since
fluctuations are small and all the molecules are
Bose-condensed. This $T=0$ limit was first studied by Eagles\cite{Eagles} and
Leggett\cite{Leggett} in the absence of a Feshbach resonance. 
As we mentioned earlier, for $T$ approaching $T_{\rm c}$, the fluctuations
associated with exciting molecules out of the condensate and
coupling to the particle-particle (Cooper-pair) channel become dominant. 
\par
These
fluctuations (rather than the breaking up of two-particle bound
states) were first included by Nozi\`eres and Schmitt-Rink
(NSR)\cite{Nozieres} to determine $T_{\rm c}$. 
In Ref. \cite{Ohashi4}, we extended this NSR approach to discuss the
superfluid phase {\it below} $T_{\rm c}$ in a {\it uniform} Fermi gas,
including a Feshbach resonance and associated molecular bosons. 
Both ${\tilde \Delta}$ and $\mu$ are obtained by solving 
Eqs. (\ref{eq.13}) and (\ref{eq.13b})
self-consistently, but now include the depletion of ${\tilde \Delta}$
through the presence of non-Bose-condensed molecules.
This procedure gives the simplest extension of the MFA-BCS
single-particle results, in that now Eqs. (\ref{eq.10}) and (\ref{eq.11}) 
involve values of $\mu$ and ${\tilde \Delta}$ which include 
(in an average way) the
effect of fluctuations around the MFA theory. 
The effect of such fluctuations is 
of considerable current interest\cite{Stajic,Perali2} and lead to what is called
the ``pseudogap" regime (for a recent review, see Ref. \cite{Levin}).
In this region, strong low-energy fluctuations in the Cooper-channel
suppress the density of states around the Fermi energy, 
which has the same effect as if there was an effective pair potential, 
even outside the
superfluid region ($T>T_{\rm c}$) where ${\tilde \Delta}(r)$ vanishes.
In a future paper, we will use this generalized NSR 
approach to include fluctuations and 
extend the results of the
present paper to finite temperatures. 
\vskip2mm
\section{Single-particle Green's functions}
\vskip2mm
In ordinary BCS superfluidity, the single-particle excitations (BCS quasiparticles) are
associated with dissociation of weakly-bound 
Cooper-pairs. In the coupled fermion-boson
model in Eq. (\ref{eq.2.6}), these Cooper pairs are replaced by
composite bosons, consisting of Cooper-pairs and molecular bosons
associated with the Feshbach resonance [${\tilde \Delta}({\bf r})
\equiv\Delta({\bf r})-g_{\rm r}\phi_M({\bf r})$]. 
 Even in the BEC regime ($2\nu<0$), Fermi excitations can still
exist\cite{Ohashi6} as well as collective modes (which form a Bose
 spectrum). As more and more fermions pair up to form bosons, the spectral
weight of the
 Fermi branch vanishes, shifting to the Bose collective branch. 
\par
Single-particle properties are most conveniently discussed in terms of
Green's functions. 
The ``diagonal" single-particle thermal Green's function 
$G_{11}({\bf r},{\bf r}',i\omega_m)$ is defined by\cite{Mahan,Sch,Fetter}
\begin{equation}
G_{11}({\bf r},{\bf r}',i\omega_m)=
-\int_0^\beta d\tau e^{i\omega_m\tau}
\Bigl\langle
T_\tau
\Bigl\{
\Psi_\uparrow({\bf r},\tau)\Psi_\uparrow^\dagger({\bf r}',0)
\Bigr\}
\Bigr\rangle,
\label{eq.2.28}
\end{equation}
where $i\omega_m$ is the fermion Matsubara 
frequency associated with the imaginary time $\tau$. 
One needs three other single-particle Green's 
functions to describe a Fermi
superfluid, as summarized by the $2\times 2$ matrix 
Green's function\cite{Sch}
\begin{eqnarray}
{\hat G}({\bf r},{\bf r}',i\omega_m)
&=&
\left(
\begin{array}{cc}
G_{11}({\bf r},{\bf r}',i\omega_m)&
G_{12}({\bf r},{\bf r}',i\omega_m)\\
G_{21}({\bf r},{\bf r}',i\omega_m)&
G_{22}({\bf r},{\bf r}',i\omega_m)\\
\end{array}
\right)
\nonumber
\\
&=&
-\int_0^\beta d\tau e^{i\omega_m\tau}
\left(
\begin{array}{cc}
\Bigl\langle
T_\tau
\Bigl\{
\Psi_\uparrow({\bf r},\tau)\Psi_\uparrow^\dagger({\bf r}',0)
\Bigr\}
\Bigr\rangle,
&
\Bigl\langle
T_\tau
\Bigl\{
\Psi_\uparrow({\bf r},\tau)\Psi_\downarrow({\bf r}',0)
\Bigr\}
\Bigr\rangle
\\
\Bigl\langle
T_\tau
\Bigl\{
\Psi_\downarrow^\dagger({\bf r},\tau)\Psi_\uparrow^\dagger({\bf r}',0)
\Bigr\}
\Bigr\rangle,
&
\Bigl\langle
T_\tau
\Bigl\{
\Psi_\downarrow^\dagger({\bf r},\tau)\Psi_\downarrow({\bf r}',0)
\Bigr\}
\Bigr\rangle
\\
\end{array}
\right).
\nonumber
\\
\label{eq.2.28b}
\end{eqnarray}
In Eq. (\ref{eq.2.28b}), 
$G_{22}$ gives the single-particle excitation spectrum
of Fermi atoms of pseudospin $\downarrow$. 
The off-diagonal components $G_{12}$ 
and $G_{21}$ arise as a direct consequence of the broken symmetry and
the presence of a condensate of Cooper pairs. Using the BdG equations
in Eq. (\ref{eq.2.5e}), one can show that these Green's functions are
related to each other as
$G_{22}({\bf r}',{\bf
  r},i\omega_m)=-G_{11}({\bf r},{\bf r}',i\omega_m)$, and  
$G^*_{21}({\bf r}',{\bf r},-i\omega_m)=G_{12}({\bf r},{\bf
  r}',i\omega_m)$.
From the definition of the Cooper-pair order parameter
$\Delta(r)=U\langle\Psi_\downarrow({\bf r})\Psi_\uparrow({\bf
  r})\rangle$, we find the important self-consistency condition
(the gap equation)
\begin{equation}
\Delta(r)={1 \over \beta}{\sum_{\omega_m}}'G_{12}
({\bf r},{\bf r},i\omega_m).
\label{eq.2.28c}
\end{equation}
\par
Using the eigenfunctions $f^F_{lmn}({\bf r})$ 
defined in Eq. (\ref{eq.2.10b}), we can write
$2\times 2$ single-particle Green's function in 
Eq. (\ref{eq.2.28b}) as
\begin{equation}
{\hat G}({\bf r},{\bf r}',i\omega_m)=\sum_{lm}
Y_{lm}({\hat \theta}){\hat g}^l(r,r',i\omega_m)Y^*_{lm}({\hat \theta}'),
\label{eq.2.29}
\end{equation}
where ${\hat g}^l(r,r',i\omega_m)$ is the $2\times 2$ Green's function for 
a given value of the 
angular momentum $l$, 
\begin{eqnarray}
{\hat g}^l(r,r',i\omega_m)
=
\sum_{j=0}^{N_l}
\Biggl[
{
\Lambda_{jl}(r)\Lambda_{jl}^\dagger(r')
\over 
i\omega_m-E^F_{jl}
}
+
{
{\bar \Lambda}_{jl}(r){\bar \Lambda}_{jl}^\dagger(r')
\over 
i\omega_m+E^F_{jl}
}
\Biggr].
\nonumber
\\
\label{eq.2.30}
\end{eqnarray}
The two-component spinor $\Lambda_{jl}(r)$ in Eq. (\ref{eq.2.30}) 
is defined in terms of 
the solutions of the BdG equations in Eq. (\ref{eq.2.21}), namely
\begin{eqnarray}
\Lambda_{jl}(r)
=
\sum_{n=0}^{N_l}
\left(
\begin{array}{c}
W^l_{n+1,j+1} \\
W^l_{{\bar N_l}+n,j+1}
\end{array}
\right)R^F_{nl}(r),~~~~{\bar \Lambda}_{jl}(r)=i\tau_2\Lambda_{jl}(r).
\label{eq.2.31}
\end{eqnarray}
Here $\tau_2$ is the Pauli matrix. 
\par
A very useful quantity is the 
Fermi single-particle excitation spectrum $N(\omega)$, also
referred to as the density-of-states (DOS). This is related to the
spectrum of single-particle Green's function,
\begin{equation}
N(\omega)=-{1 \over \pi}
\int d{\bf r}
Im
\Bigl[
G_{11}({\bf r},{\bf r},i\omega_m\to\omega+i\delta)
\Bigr].
\label{eq.2.27}
\end{equation}
Substituting Eqs. (\ref{eq.2.29}) and (\ref{eq.2.30}) into
Eq. (\ref{eq.2.27}), this {\it spatially-averaged} density-of-states 
(per pseudospin) is given by ($E^F_{jl}>0$)
\begin{eqnarray}
N(\omega)=
\sum_{l}(2l+1)
\sum_{n,j}
|W^l_{n+1,j+1}|^2\delta(\omega-E^F_{jl})~~~(\omega\ge0).
\label{eq.2.32}
\end{eqnarray}
It is also useful to introduce the {\it local} density of states (LDOS),
defined as
\begin{eqnarray}
N(\omega,r)
&\equiv&
-{1 \over \pi}
Im
\Bigl[
G_{11}({\bf r},{\bf r},i\omega_m\to\omega+i\delta)
\Bigr]
\nonumber
\\
&=&
\sum_{nn'l}
{2l+1 \over 4\pi}
R_{nl}^F(r)R_{n'l}^F(r)
\sum_j
W_{n+1,j+1}^lW_{n'+1,j+1}^l
\delta(\omega-E^F_{jl}).
\label{eq.2.33}
\end{eqnarray}
The local density of states LDOS in Eq. (\ref{eq.2.33}) is simply
related to the spatially-averaged total DOS in
Eq. (\ref{eq.2.32}) by
\begin{equation}
N(\omega)\equiv \int_0^\infty 4\pi r^2drN(\omega,r).
\label{eq.2.34}
\end{equation}
\par
The gap equation (\ref{eq.2.23}) is easily obtained from the 
present Green's function formalism. 
When we use the off-diagonal (1,2)-component of Eq. (\ref{eq.2.30}) in
Eq. (\ref{eq.2.28c}), we obtain
\begin{eqnarray}
\Delta(r)=-U{\sum_{nn'l}}'
{2l+1 \over 4\pi}
R_{nl}^F(r)R_{n'l}^F(r)
\sum_{j=0}^{N_l}
W^l_{{\bar N}_l+n,j+1}
W^l_{n'+1,j+1}.
\label{eq.2.31b}
\end{eqnarray}
Once one has $\Delta(r)$, one can calculate the coefficients $\alpha_n$
in Eq. (\ref{eq.2.18}) and then find $\phi_M(r)$. By this procedure, one finally
obtains the value of ${\tilde \Delta}(r)$, with the total number
equation in Eq. (\ref{eq.2.26}) determining the self-consistent
values.
As with the usual BdG equations in a uniform BCS model, there is 
a relation between eigenfunctions for 
$E^F_{nl}\ge 0$ and $E_{nl}^F\le 0$, which is given by\cite{BdG}
\begin{eqnarray}
\sum_n
\left(
\begin{array}{c}
W_{{\bar N}_l+n,j+1}^l\\
-W_{n+1,j+1}^l\\
\end{array}
\right)
R_{nl}^F(r)
=
\sum_n
\left(
\begin{array}{c}
W_{n+1,{\bar N}_l+j}^l\\
W_{{\bar N}_l+n,{\bar N}_l+j}^l\\
\end{array}
\right)
R_{nl}^F(r).
\label{eq.2.31c}
\end{eqnarray}
Substituting Eq. (\ref{eq.2.31c}) into Eq. (\ref{eq.2.31b}), 
we find that the gap equation (\ref{eq.2.23}) for the Cooper-pair order parameter
$\Delta(r)$ is reproduced.
\vskip5mm
\section{Scattering lengths, gap equation and strength of Feshbach resonances}
\vskip2mm
The Hamiltonian in Eq. (\ref{eq.2.1}) involves the bare energies $U$, $g_{\rm r}$ 
and $2\nu$. As usual in dealing with ultracold atomic gases, it is convenient to
work in terms of 
renormalized interaction energies which incorporate the effect of high energy
processes. This procedure naturally leads to the two-body scattering length $a_s^{2b}$
which describes the effective interaction between low energy atoms, even in the
case of a Feshbach resonance.
The two-body scattering length $a_s^{2b}$ 
can be measured directly in a variety of ways.
\par
In this section, we briefly review the standard theory for renormalized low
energy parameters (see, for example, Sec. IV A of Ref. \cite{Ohashi3} and Refs. \cite{Milstein,Holland1,Randeria}).
We also point out that in the presence of a Feshbach resonance, the
self-consistent gap equation which determines the order parameter naturally
introduce a {\it different} 
$s$-wave scattering length $a_s$, which has a crucial dependence on
the Fermi chemical potential $\mu$.
As a result, in dealing with the BCS-BEC crossover in Fermi superfluids,
it seems most natural to treat $a_s$ as the control parameter, rather than $a_s^{2b}$.
\par
There is a second reason which makes it useful to discuss properties in
crossover region as a function of $a_s$.
Before doing so, we discuss the parameters we use in this paper.
We take the total number of atoms to be $N=10,912$ 
$(=N_\uparrow+N_\downarrow=2N_\uparrow)$.
In a non-interacting Fermi gas, this corresponds to filling atoms (per spin) up to
$E=31.5\hbar\omega_0$ ($\equiv\varepsilon_{\rm F}$), 
where $\omega_0$ is the trap frequency.
As the unit of length, we use the 
Thomas-Fermi radius
$R_{\rm F}\equiv\sqrt{2\varepsilon_{\rm F}/m\omega_0^2}$ for a 
free Fermi gas in the trap.
We take the Feshbach coupling constant $g_{\rm r}=0.06\omega_0$ 
(${\bar g}_{\rm r}\equiv g_{\rm r}(\sqrt{N/R_{\rm F}^3})
=0.2\varepsilon_{\rm F}$). The non-resonant pairing 
interaction values we use are either $U=0.001\omega_0$ 
(${\bar U}\equiv U(N/R_{\rm F}^3)=0.35\varepsilon_{\rm F}$) or $0.0015\omega_0$ (${\bar
  U}=0.52\varepsilon_{\rm F}$). 
For the high-energy cutoff, we take $\omega_c=161.5\hbar\omega_0$ 
($\gg\varepsilon_{\rm F}$). 
\par
The explicit calculations presented in this paper are for what is called
a narrow Feshbach resonance, while all current ultracold Fermi gas experiments
are done using broad Feshbach resonances.
However, in a uniform Fermi gas, several quantities are found to have values 
very similar if viewed as a function of $a_s$, for both weak (${\bar g}_{\rm r}<\varepsilon_{\rm F}$) and strong (${\bar g}_{\rm r}>\varepsilon_{\rm F}$) Feshbach resonances.
At the present time, numerical calculations are only able to deal with a narrow
Feshbach resonance in the case of a trapped Fermi gas. This is due to the fact that
one is limited to dealing with a finite number of excited states in a trap 
and this makes it difficult to deal with a broad Feshbach resonance, which couples molecules to Fermi atoms in very high-energy eigenstates.
This approximate independence of the strength of the Feshbach resonance (for a given
value of $a_s$) is thus very useful.
\par
We first recall the standard case of two Fermi atoms 
interacting in a vacuum\cite{Fetter}
with a bare interaction denoted by $-U$.
In this case, the effective low-energy ($\omega=0$) renormalized $s$-wave scattering
length is given by
\begin{equation}
-{4\pi a_s^{2b} \over m}=
{U \over 1-U\sum_{[0,\omega_c]}{1 \over 2\varepsilon_{\bf p}}}
\equiv U^R.
\label{eq.bcs4}
\end{equation}
Effectively the renormalized interaction $U^R$ incorporates 
all high energy scattering
processes. In the absence of a Feshbach resonance ($g_{\rm r}=0$) 
and in a uniform gas,
the gap equation (\ref{eq.13}) 
can be written in terms of this renormalized interaction as follows:
\begin{eqnarray}
1=U^R\sum_{[0,\omega_c]}
\Bigl[
{1 \over 2E_{\bf p}}\tanh{E_{\bf p} \over 2T}-{1 \over 2\varepsilon_{\bf p}}
\Bigr].
\label{eq.bcs3}
\end{eqnarray}
Here we have written the gap equation at finite temperatures. 
The summation in Eq. (\ref{eq.bcs3}) converges, so we can safely 
takes $\omega_c \to\infty$. 
This gives a cutoff-independent gap equation as a function of the renormalized interaction $U^R$ or, equivalently, 
the two-body scattering length $a_s^{2b}$ defined in Eq. (\ref{eq.bcs4}).
As first discussed by Leggett\cite{Leggett} and Randeria\cite{Randeria}, it is
convenient to discuss the BCS-BEC crossover region in terms of the dimensionless
parameter $(k_{\rm F}a^{2b}_s)^{-1}$. As the bare attractive interaction $U$ increases,
$(k_{\rm F}a^{2b}_s)^{-1}$ goes from $-\infty$ (BCS) to $\infty$ (BEC).
\par
In the case of a Feshbach resonance, one can introduce a renormalized $s$-wave scattering length describing low energy atoms which is the analogue of Eq. (\ref{eq.bcs4}), 
namely,
\begin{equation}
-{4\pi a^{2b}_s \over m}
\equiv
{U_{\rm eff}^{2b} \over 1-U_{\rm eff}^{2b}\sum_{[0,\omega_c]}
{1 \over 2\varepsilon_{\bf p}}},
\label{eq.bcs4bb}
\end{equation}
where the bare Feshbach resonance is described by
\begin{equation}
U_{\rm eff}^{2b}\equiv U+{g_{\rm r}^2 \over 2\nu}.
\label{eq.bcs01}
\end{equation}
Once can rewrite Eq. (\ref{eq.bcs4bb}) in the form
\begin{equation}
-{4\pi a_s^{2b} \over m}=U^R+{(g^R_{\rm r})^2 \over 2\nu^R}, 
\label{eq.bcs4p}
\end{equation}
which now involves the renormalized parameters for $U$, $g_{\rm r}$, and $2\nu$
(for details, see, for example, Sec. IV A of Ref. \cite{Ohashi3}). These
low energy renormalized parameters can be directly measured.
\par
However, in contrast to Eq. (\ref{eq.bcs3}), the cutoff-independent energy gap
equation in the presence of a Feshbach resonance in {\it not} given simply in terms of
the usual 
two-body $s$-wave 
scattering length as defined in Eqs. (\ref{eq.bcs4bb}) and (\ref{eq.bcs4p}).
Instead, one finds\cite{Ohashi1,Ohashi3,Ohashi4,Milstein}
\begin{eqnarray}
1=U^R_{\rm eff}\sum_{[0,\omega_c\to\infty]}
\Bigl[
{1 \over 2E_{\bf p}}\tanh{E_{\bf p} \over 2T}-{1 \over 2\varepsilon_{\bf p}}
\Bigr],
\label{eq.bcs3b}
\end{eqnarray}
where 
\begin{eqnarray}
U_{\rm eff}^R
&\equiv&
{U_{\rm eff} \over 1-U_{\rm eff}\sum_{[0,\omega_c]}{1 \over 2\varepsilon_{\bf p}}}
\nonumber
\\
&\equiv&
U^R+
{(g_{\rm r}^R)^2
\over
2\nu^R-2\mu
}
\nonumber
\\
&\equiv&
-{4\pi a_s \over m}.
\label{eq.bcs02}
\end{eqnarray}
$U_{\rm eff}$ is defined in Eq. (\ref{eq.2.5d}).
We first restrict the following 
discussion to the case of a uniform Fermi superfluid.
Comparing this new $s$-wave scattering length $a_s$ involved in the gap equation (\ref{eq.bcs3b}) with the the two-body scattering length $a_s^{2b}$ defined by 
Eqs. (\ref{eq.bcs4bb}) and (\ref{eq.bcs4p}), we see that $a_s$ depends crucially
on the Fermi chemical potential $\mu$. This in turn is a strong function
of the threshold energy $2\nu$ and hence will change value in the BCS-BEC crossover.
One may think of $a_s$ as including many-body effects related to the coupling 
of the fermions to molecular bosons\cite{notenum3}. 
We note in passing that it is wrong to
simply use $a_s^{2b}$ [given in Eq. (\ref{eq.bcs4p})] in the gap equation
(\ref{eq.bcs3b}) in the general case.
\par
The fact that the cutoff-free gap equation (\ref{eq.bcs3b}) {\it only} depends on $a_s$, {\it irrespective of detailed values of the non-resonant interaction $U$ and the Feshbach resonance strength $g_{\rm r}$}, is very useful.
In particular, a broad Feshbach resonance (${\bar g}_{\rm r}>\varepsilon_{\rm F}$) and a narrow Feshbach resonance (${\bar g}_{\rm r}<\varepsilon_{\rm r}$) give approximately the same values for $T_{\rm c}$ and ${\tilde \Delta}$ 
{\it for the same value of $a_s$}. 
As an example, we show in Fig. \ref{fig1}(a) the phase transition temperature 
$T_{\rm c}$ in the BCS-BEC crossover for a {\it uniform} gas for a broad and a 
narrow Feshbach resonances. In a broad 
Feshbach resonance, the BCS-BEC crossover occurs at the value
of $\nu\gg\varepsilon_{\rm F}$. In the case of a narrow Feshbach 
resonance, the crossover occurs at $\nu\sim\varepsilon_{\rm F}$.
However, when we express $T_{\rm c}$ as a function of  $(k_{\rm F}a_s)^{-1}$, 
both cases give almost the same result over the whole
BCS-BEC crossover region, as shown in Fig. \ref{fig1}(a)\cite{notenum4}.
In addition, Fig. \ref{fig1}(b) shows that the change 
from a gas of Fermi atoms to a molecular Bose gas in the BCS-BEC
crossover is also almost the same for both large and small values of 
${\bar g}_{\rm r}$ when viewed as a function of $(k_{\rm F}a_s)^{-1}$. 
The region of $(k_{\rm F}a_s)^{-1}>0$ can thus be regarded
as a BEC of molecular Bosons for both broad and narrow Feshbach
resonances. 
Using $a_s$ in Eq. (\ref{eq.bcs02}),
we find the same crossover physics irrespective of the 
width of the Feshbach resonance. 
This is useful since as we noted earlier, 
our numerical calculations are limited to ${\bar g}_r\lesssim\varepsilon_{\rm F}$, while
recent experiments deal with a broad Feshbach resonance.
We also note that in the case of a broad Feshbach resonance, the crossover
occurs  at a value of $\nu\gg\varepsilon_{\rm F}>\mu$. In this region,
we can omit the chemical potential in Eq. (\ref{eq.bcs02}) and then one finds
$a_s\simeq a_s^{2b}$, the usual renormalized low-energy 
two-body scattering length
which can measured directly.
\par
In contrast to $T_{\rm c}$, the character of the 
Feshbach resonance (broad or narrow resonance) does
show up somewhat 
when we consider the number of molecules $N_{\rm M}$ in Fig. \ref{fig1}(b).
With increasing $(k_{\rm F}a_s)^{-1}$ from the BCS side, 
Fig. \ref{fig2}(b) shows that the number of 
Feshbach molecules with a finite lifetime increases in the case of a
{\it narrow} Feshbach 
resonance. Stable molecules 
($N_{\rm B}^{\gamma=0}$) become dominant 
in the region of negative chemical potential, $\mu<0$ (see Fig. \ref{fig2}(a)). 
Although stable
Cooper-pairs given by the $N_{\rm C}^{\gamma=0}$ curve 
also exist in this region,
we see that $N_{\rm C}^{\gamma=0}\ll N_{\rm B}^{\gamma=0}$. Thus, in the
case of a narrow
Feshbach resonance (small ${\bar g}_{\rm r}$), superfluidity in the region of
$\mu<0$ is largely associated with a BEC of stable Feshbach
molecules. 
\par
On the other hand, in the case of a {\it broad} Feshbach resonance, 
the number of Cooper-pairs with a finite lifetime 
($N_{\rm C}^{\rm sc}$) first rapidly 
increases as one increases $(k_{\rm F}a_s)^{-1}$ 
[see Fig. \ref{fig2}(c)]. When $\mu<0$,
stable long lived Cooper-pairs ($N_{\rm C}^{\gamma=0}$) become the dominant
bound states.
Below $T_{\rm c}$, the tightly-bound molecules 
around $\mu\lesssim 0$ are now Cooper-pairs.
As shown in Fig. \ref{fig2}(c), the Feshbach molecules 
($N_{\rm B}^{\gamma=0}$) eventually
become dominant deep inside the BEC
regime (defined by $(k_{\rm F}a_s)^{-1}\sim 1$).
The difference between the two cases is due the fact that, in a broad
Feshbach resonance, the BCS-BEC crossover occurs around 
$\nu\gg\varepsilon_{\rm F}$, where the formation/dissociation
 of Feshbach molecules only
appears as a virtual process. In contrast, the crossover region
is located in the region $\nu\lesssim\varepsilon_{\rm F}$ in the case of 
a narrow Feshbach resonance,
where a large number of Feshbach molecules can form as quasi-stable entities.
\par
The preceding discussion was limited to a {\it uniform} Fermi superfluid.
A discussion on $T_{\rm c}$ and the composite order parameter in terms of
the scattering length $a_s$ is also possible in a trapped gas, if we use
the LDA\cite{Ohashi3,Chiofalo}. In this case, the 
spatially-dependent order parameter ${\tilde \Delta}(r)$ is 
self-consistently determined by the 
cutoff-dependent gap equation, given by 
\begin{equation}
1=U_{\rm eff}(r)\sum_{[0,\omega_c]}{1 \over 2E_{\bf p}(r)}
\tanh {E_{\bf p}(r) \over 2T}.
\label{eq.bcs1bbb}
\end{equation}
Here $E_{\rm p}(r)$ is the energy of a BCS excitation 
with density and other quantities evaluated locally at position $r$. This 
is given by Eq. (\ref{eq.11}), with $\mu$ being replaced by 
$\mu(r)\equiv \mu-m\omega_0^2r^2/2$. The bare pairing interaction strength 
at $r$ in Eq. (\ref{eq.bcs1bbb}) is defined as 
\begin{equation}
U_{\rm eff}(r)\equiv
U+{g_{\rm r}^2 \over 2\nu+(3/2)\omega_0-2\mu(r)},
\label{eq.4.1a}
\end{equation}
where we have included the zero point energy $(3/2)\omega_0$ for later discussions. 
From our preceding analysis, the LDA gap equation (\ref{eq.bcs1bbb}) can be written
in a cutoff-independent form, but it now involves the renormalized pairing 
interaction given by
\begin{equation}
U_{\rm eff}^R(r)\equiv
{U_{\rm eff}(r) 
\over 1-U_{\rm eff}(r)\sum_{[0,\omega_c]}{1 \over 2\varepsilon_{\bf p}}}.
\label{eq.4.1b}
\end{equation}
In particular, since $\mu(r=0)=\mu$, $U^R_{\rm eff}(r=0)$ equals $U^R_{\rm eff}$ given in Eq (\ref{eq.bcs02}) [apart from the zero point energy 
term $(3/2)\omega_0$ in Eq. (\ref{eq.4.1a})]. Thus, recalling that $T_{\rm c}$ in the LDA\cite{Ohashi3} is determined by the gap equation (\ref{eq.bcs1bbb}) at $r=0$, we conclude that the crossover behavior of $T_{\rm c}$ and ${\tilde \Delta}(r=0)$ at the center of the trap can be described (as in a uniform Fermi gas) as functions of the renormalized scattering length $a_s$ defined in Eq. (\ref{eq.bcs02}). Indeed, Fig. \ref{fig3}(a) shows that both narrow and broad Feshbach resonance cases give almost the same $T_{\rm c}$ as a function of $(k_{\rm F}a_s)^{-1}$ in the whole BCS-BEC crossover regime.
\par
As in a uniform gas discussed above, in a trap 
some quantities will depend on the strength of the Feshbach resonance.
The LDA gap equation (\ref{eq.bcs1bbb}) involves a position-dependent $a_s(r)$ 
because of the spatial dependence of $\mu(r)$.
In a broad Feshbach resonance 
(${\bar g}_{\rm r}\gg\varepsilon_{\rm F}$), the crossover occurs at $\nu\gg\varepsilon_{\rm F}$, so that $\mu(r)$ ($\sim\varepsilon_{\rm F}$) can be neglected in this regime. On the other hand, $U_{\rm eff}(r)$ decreases as we go from the center of the trap in the case of a narrow Feshbach resonance (${\bar g}_{\rm r}<\varepsilon_{\rm F}$), since now the crossover occurs at $\nu\sim\varepsilon_{\rm F}$ and hence $\mu(r)$ cannot be ignored in $U_{\rm eff}(r)$.
This explains why the total number of molecules in slightly smaller in the case of a narrow Feshbach resonance, as shown in Fig. \ref{fig3}(b) 
\par
In this paper, we present results as a functions of 
the renormalized uniform gas scattering length $a_s$ in 
Eq. (\ref{eq.bcs02}), with the result that the theory is approximately cutoff-free 
within LDA. The dependence on $(k_{\rm F}a_s)^{-1}$ also gives a description 
for Fermi superfluids around the center 
of the trap, irrespective of the width of the Feshbach resonance. 
We refer to the region of $(k_Fa_s)^{-1}<0$ as the BCS regime
and the region of $(k_Fa_s)^{-1}>0$ as the BEC regime. 
We note that the ``real" or bare pairing interaction in the gap equation is {\it not} 
$4\pi a_s/m$, but $-U_{\rm eff}$ defined in Eq. (\ref{eq.2.5d}). This is always 
{\it attractive} even in 
the positive (BEC) region of $a_s>0$, as we show in Fig. \ref{fig4}(a).
In Fig. \ref{fig4}(b),
we give the relation between $a_s$ and $\nu$, for the parameters $g_{\rm r}$ and $U$
used in this paper.
In the numerical results discussed in this paper, for the convenience of the reader, 
we give both the value of $\nu$ as well as the corresponding value of $(k_{\rm F}a_s)^{-1}$.
\par
\vskip5mm
\section{Equilibrium properties in the crossover region}
\vskip2mm
Figure \ref{fig5} shows the calculated atomic density profile at different
places in the BCS-BEC crossover
region at $T=0$. 
In the BCS regime shown in Fig. \ref{fig5}(a), the number of Fermi atoms is
much larger than the number of (Feshbach) molecules.
Although these atoms form 
Cooper-pair bosons, the Pauli exclusion principle
is still relevant. As a result, the Fermi gas spreads out to a distance of the
order of the Thomas-Fermi
radius $R_{\rm F}$, just as in the case of a trapped non-interacting Fermi gas. 
We recall that in contrast to bosons, interactions have a negligible
effect on the density profile of a trapped Fermi gas in the normal
phase. 
The relative number of Cooper-paired fermions [$n_{\rm F}(r)$] and 
bound (Feshbach) dimers [$n_{\rm M}(r)$]
continuously changes as the threshold energy $2\nu$ is
lowered (or $(k_{\rm F}a_s)^{-1}$ increases).
On the BCS side of the crossover regime shown in Fig. \ref{fig5}(b), the molecular
density $n_{\rm M}(r)$ has become very large in the center of 
the trap. This shows that the Pauli exclusion principle, as expected, 
is less important in this
region where Bose molecules start to dominate.
However, the Fermi atoms still dominate for $r\gesim 0.2R_{\rm F}$. 
This feature persists 
on the BEC side of crossover regime shown in Fig. \ref{fig5}(c),
where $n_{\rm F}(r)$ is seen to be dominant 
for $r\gesim 0.3R_{\rm F}$. The tail of the density profile consisting
of unpaired atoms finally disappears in Fig. \ref{fig5}(d), where
almost all the atoms have formed Feshbach molecules. 
\par
Figure \ref{fig6} shows how the Fermi chemical potential $\mu$ changes in the BCS-BEC
crossover. As one approaches the BEC regime, $\mu$ decreases from the 
weak-coupling (BCS) result given by 
$\mu\simeq \varepsilon_{\rm F}~(=31.5\omega_0$ in our case).
In the BEC regime, the molecular
Bose chemical potential $\mu_M=2\mu$ approaches the lowest molecular
energy, given by $2\nu+(3/2)\omega_0$, as shown in the inset in Fig. \ref{fig6}. 
We note
that this limiting case of $\mu_M=2\nu+(3/2)\omega_0$ is just the condition 
for the BEC of a non-interacting Bose gas in a harmonic trap.
\par
In Fig. \ref{fig6}, 
the Fermi chemical potential is seen to go through
 zero at $(k_{\rm F}a_s)^{-1}\simeq 0.65$, which corresponds to $\nu=0$.
To understand this, let us consider the case of a {\it uniform} gas. In this case,
the single-particle excitation spectrum is given by the 
usual BCS expression (but now with 
the composite order parameter ${\tilde \Delta}$)
\cite{Ohashi4,Leggett,Randeria,Ohashi6},
\begin{equation}
E_{\bf p}=\sqrt{(\varepsilon_{\bf p}-\mu)^2+{\tilde \Delta}^2}.
\label{eq.3.2}
\end{equation}
In the weak-coupling BCS regime, where $\mu$ is positive, the
energy gap of the BCS
excitations at the Fermi energy 
is equal to $|{\tilde \Delta}|$. However, this relation
is no longer valid when $\mu$ is negative, as in the strong-coupling
crossover regime\cite{Leggett,Melo}. In this regime, the energy gap 
is given by
\begin{equation}
E_g=\sqrt{\mu^2+{\tilde \Delta}^2}.
\label{eq.3.2b}
\end{equation}
In the BEC limit ($\nu\to-\infty$), 
the composite
order parameter is described by a pure BEC condensate of molecules, given by
${\tilde \Delta}\to g_{\rm r}\sqrt{N/2}$. In the same limit, the chemical 
potential approaches
the threshold energy $\mu\simeq \nu\to-\infty$. 
Thus, the excitation gap $E_g$ given by Eq. (\ref{eq.3.2b}) 
approaches $|\nu|$. This makes sense, since $2\nu~(<0)$ is 
the lowest molecular energy, and hence $2|\nu|$ is the 
excitation energy to dissociate a molecule into {\it two} Fermi atoms.
\par  
In a trapped gas, within the LDA, the chemical potential $\mu$ and 
the order parameter ${\tilde \Delta}$ in 
Eq. (\ref{eq.3.2b}) are replaced by the position-dependent ones,
$\mu(r)\equiv \mu-m\omega_0^2r^2/2$ and ${\tilde \Delta}(r)$, respectively. 
In the extreme BEC limit
($\nu\to-\infty$), the lowest single-particle excitation
energy is given as the energy to dissociate a molecule into
two Fermi atoms and put them at the lowest (unpaired) fermion state.
In the LDA, this lowest state is at $r=0$, because
the trap potential has its minimum at the center [$V^{\rm F}_{\rm trap}(r=0)=0$]. 
This again leads to
$E_g\sim|\mu|\sim|\nu|$ in the BEC limit (when ${\tilde \Delta}(0)\ll|\mu|$).
As a result, we find that 
$(k_{\rm F}a_s)^{-1}\simeq 0.65$ (at which $\mu=0$) gives a 
characteristic
scattering length which separates 
the region dominated by weakly-bound Cooper-pairs
($\mu>0$) and the region dominated by tightly-bound molecules ($\mu<0$).
Equivalently, this boundary occurs at $\nu=0$.
\par
Figure \ref{fig7} shows the profile of composite order parameter ${\tilde \Delta}(r)$. 
As expected from Fig. \ref{fig5}, ${\tilde \Delta}(r)$ is dominated by the Cooper-pair
component $\Delta(r)$ in the BCS regime shown in panel \ref{fig7}(a). This calculated
profile agrees with previous results in the BCS limit, as obtained by
Bruun and co-workers\cite{Bruun0}.
However, in the presence of a Feshbach resonance, even in the
BCS regime, the molecular condensate $\phi_M(r)$
is found to be finite everywhere $\Delta(r)$ is finite, 
due to the identity given in Eq. (\ref{eq.2.4}).
In an ideal Bose gas BEC, $\phi_M(r)$ is simply proportional 
to the ground state wavefunction of 
the harmonic potential given by Eq. (\ref{eq.2.18b}). However, since the
molecular bosons are strongly hybridized with the Fermi atoms by the Feshbach 
resonance, in the BCS regime, we find that $\phi_M(r)$ is no longer
given by the ground state of the harmonic potential as in 
Eq. (\ref{eq.2.18b}). Indeed, as shown in Fig. \ref{fig8}(a), $\phi_M(r)$
in this regime is given by a superposition of excited molecular 
states ($n\ge 1$), as indicated in Eq. (\ref{eq.2.17}).
The molecular condensate $\phi_M(r)$ continuously
increases in magnitude 
as one approaches the BEC regime, as shown in Figs. \ref{fig7}(b) and (c). 
At the same time, Fig. \ref{fig8} shows that the contribution of excited states
($n\ge 1$) decreases, indicating that $\phi_M(r)$
approaches the ideal BEC described by Eq. (\ref{eq.2.18b}).
However, because of the identity in Eq. (\ref{eq.2.4}), 
the Cooper-pair component $\Delta(r)$ remains finite even in the BEC regime.
Indeed, as shown in Fig. \ref{fig7}(d), 
$\Delta(r=0)\simeq 15\omega_0$ when $\nu=0$.
\par
In Fig. \ref{fig9}, we compare the total atomic 
density profile $n(r)=2n_{\rm M}(r)+n_{\rm F}(r)$ 
with the profile of the
composite order parameter ${\tilde \Delta}(r)$, which clearly
shows that the composite order parameter is always 
finite in the region where $n(r)$ is
finite. In addition, we expect that $n(r)\simeq |\phi_M(r)|^2$ in the BEC regime.
This is confirmed by the calculated values shown 
in the inset in Fig. \ref{fig9}(b), which show that 
$n(r)\simeq |{\tilde \Delta}(r)|^2$. 
\par
We note that when the LDA is used in calculating the density profile as well as 
the order parameter at $T=0$, it has been shown that 
the LDA is a good approximation in the BCS regime\cite{Bruun0}. In
contrast, the shrinkage of the density profile $n(r)$ shown in Fig. \ref{fig7}, 
as well as the
profile of the composite order parameter ${\tilde \Delta}(r)$, is
poorly overestimated by the LDA\cite{Chiofalo} as we enter the BEC regime. 
This is because the LDA
underestimates the kinetic energy, which results in the density profile of atoms
spreading out more.
\par
At $T=0$, such an LDA calculation correctly 
predicts that the resulting composite order
parameter is finite in all regions where $n(r)$ is finite. This reflects the
fact that the Fermi surface is unstable against an infinitesimally
weak attractive interaction at $T=0$, which leads to a superfluid phase
transition everywhere (as long as $n(r)$ is positive) 
in a trap in the LDA.  
However, this situation is no
longer satisfied at finite temperatures. As we have discussed in Ref. \cite{Ohashi3},
only the center of the trap ($r=0$) is in the superfluid phase {\it
  just below} $T_{\rm c}$ in the LDA, with this superfluid region becoming wider
with decreasing temperatures. Thus, for $0<T<T_{\rm c}$, an LDA
calculation predicts a spatial region where the order parameter 
vanishes, even though the particle density is still finite. (This feature 
is shown in Fig. 1 of Ref. \cite{Torma}.) 
However, the presence of such a ``two-phase" trapped Fermi gas
(superfluid at the center and normal phase at the edge of the trap) 
is clearly an artifact of the LDA. In fact, 
the entire gas is in the superfluid
phase below $T_{\rm c}$, with ${\tilde \Delta}(r)$
finite everywhere where $n(r)$ is finite. 
\vskip5mm
\section{Single-particle excitations in the crossover region}
\vskip2mm
Figure \ref{fig10} shows single-particle density-of-states $N(\omega)$ given in
Eq. (\ref{eq.2.32}). 
One sees that the excitation spectrum in a Fermi superfluid has a finite
energy gap ($\equiv E_{\rm g}$). Although the pairing interaction
becomes stronger with decreasing threshold energy $2\nu$, 
the magnitude of $E_{\rm g}\sim 1.2\omega_0$ 
is almost the same in panels 
(a) and (b) of Fig. \ref{fig10}. 
We also note that there is characteristic sharp increase or shoulder around $E_g$.
Recalling the case of a {\it uniform} BCS superfluid 
(where the energy gap is equal to $|{\tilde \Delta}|$),
one has the well-known result
\begin{equation}
N(\omega)=\rho(\varepsilon_{\rm F})
{\omega \over \sqrt{\omega^2-{\tilde \Delta}^2}}
\Theta(\omega-|{\tilde \Delta}|),
\label{eq.y1}
\end{equation}
where $\Theta(x)$ is the step
function. The normal-state density of states is given by 
$\rho(\omega)\equiv{m\sqrt{2m\omega} \over 2\pi^2}$. 
[We have approximated this by the value at the
Fermi energy, $\omega=\varepsilon_{\rm F}$, in Eq. (\ref{eq.y1}). This is a good
approximation in the BCS limit. The more correct expression for
$N(\omega)$, including the effect of the energy-dependence of 
$\rho(\omega)$, is given in Ref. \cite{Ohashi4}.] 
Equation (\ref{eq.y1}) predicts that $N(\omega)$ 
is singular at $\omega=|{\tilde \Delta}|$. 
The sudden increase in $N(\omega)$ at $\omega=E_g$ in the {\it trapped} gas 
shown in Figs. \ref{fig10}(a) and (b)
may be viewed as the remanent of this singular behavior 
of the single-particle excitations in the uniform BCS case.
\par
Figure \ref{fig10}(c) shows the change in the density of states
$N(\omega)$ as we enter into the BEC region ($\nu<0$). One finds that
the energy gap now sharply increases ($E_g\simeq 5\omega_0$) and 
the sudden jump in the magnitude of $N(\omega)$ at the energy gap $E_g$ is absent.
\par
In Fig. \ref{fig11}, we plot the single-particle excitation gap $E_g$
in the BCS-BEC crossover region\cite{notePRL}. 
This figure shows that $E_g$ rapidly increases when the chemical
potential $\mu$ becomes negative [$(k_{\rm F}a_s)^{-1}>0.65$]. In this region, 
$2E_{\rm g}$ is the dissociation energy of a molecule. This
energy $2E_g \simeq |\mu_{\rm M}|\simeq 2|\nu|$ becomes large 
as we decrease the threshold energy $2\nu\to-\infty$. 
Indeed, we find that $E_{\rm g}$ approaches 
$|\mu|$ (given by the dashed line in Fig. \ref{fig11}) 
when we take $\mu\to-\infty$, which is consistent 
with Eq. (\ref{eq.3.2b}) valid for a uniform gas. We also see
from Fig. \ref{fig12} that the number of molecules becomes much larger than the number
of Fermi atoms
for $\nu<0$ or $(k_{\rm F}a_s)^{-1}>0.65$, as expected. 
\par
One can understand (in the case of a uniform gas) why the ``coherence
peak" (i.e., the sudden jump) in $N(\omega)$ at $E_g$ is absent in the
crossover region shown in Fig. \ref{fig10}(c). 
In the BEC region one finds (see Eq. (5.4) of Ref. \cite{Ohashi4})
\begin{equation}
N(\omega)=
{m\sqrt{2m} \over 4\pi^2}
\Bigl(
1+{\omega \over \sqrt{\omega^2-{\tilde \Delta}^2}}
\Bigr)
\Bigl(
\sqrt{\omega^2-{\tilde \Delta}^2}+\mu
\Bigr)^{1/2}
\Theta(\omega-\sqrt{\mu^2+{\tilde \Delta}^2}),
\label{eq.y2}
\end{equation}
where the factor $(\sqrt{\omega^2-{\tilde \Delta}^2}+\mu)^{1/2}$ comes
from the normal state density-of-states $\rho(\omega)\propto\sqrt{\omega}$. 
When $\mu$ is
negative, the threshold energy of the $N(\omega)$ is determined not by
the factor $\sqrt{\omega^2-{\tilde \Delta}^2}$, but by the threshold
energy of the normal DOS $\rho(\omega)$. This leads to the step function in
Eq. (\ref{eq.y2}). As a result, the density-of-states
$N(\omega)$ of a Fermi superfluid is finite only for $\omega\ge\sqrt{\mu^2+{\tilde
    \Delta}^2}$. The expected coherence peak at $\omega=|{\tilde
  \Delta}|<\sqrt{\mu^2+{\tilde \Delta}^2}$ actually occurs in a region where
$N(\omega)$ vanishes.
\par
When we compare the single-particle excitation gap $E_g$ for $\mu>0$ 
in Fig. \ref{fig11} with
the magnitude ${\tilde \Delta}(r=0)$ of the composite order 
parameter shown in Fig. \ref{fig13}, 
we see that $E_g\ll|{\tilde \Delta}(0)|\sim\varepsilon_{\rm F}$ 
in the crossover region. 
Since the excitation gap $E_g$ equals 
the order parameter (when $\mu>0$) in a uniform gas, this difference clearly
originates from the
inhomogeneity of a superfluid Fermi gas in a trap. To understand 
the origin of this very small value of the single-particle excitation gap 
$E_g$ in a {\it trapped} superfluid Fermi gas,
it is very useful to compare $N(\omega)$ and the
 local density of states $N(\omega,{\bf r})$ given by the expression in 
Eq. (\ref{eq.2.33}). 
Figure \ref{fig14}(b) 
shows that no low energy spectral weight comes from the center of the trap.
In fact, panel (d) clearly shows that the 
low-energy spectral weight in $N(\omega)$ shown in panel (a) mainly
comes from the region $r\sim0.6R_{\rm F}$, which is close to the bottom of the
{\it effective potential well} composed of 
the (composite) pair potential ${\tilde \Delta}(r)$ and
the harmonic trap measured from the chemical potential $\mu$, given by 
$V_{\rm trap}^{\rm eff}(r)\equiv(V_{\rm trap}^F(r)-\mu)\Theta(V_{\rm trap}^F(r)-\mu)$
[see the inset of Fig. 14(b)]. 
Strictly speaking, this effective potential is a combination of an
ordinary (diagonal) and an anomalous (off-diagonal) potential
in the BdG equations.
This situation is analogous to the boundary
problem in superconductivity, as schematically 
indicated in Fig. \ref{fig15}. As first discussed
by de Gennes and Saint-James\cite{dG2} in 
the context of superconductivity,
``Andreev" bound states 
can appear well below the bulk energy gap
($\sim{\tilde \Delta}$), around the minimum of the effective potential well. 
Thus $E_g$ may be viewed as the energy of the lowest Andreev bound state formed
in the effective potential well [$={\tilde \Delta}(r)+V_{\rm trap}^{\rm eff}(r)$].
These are also called the ``in-gap" states\cite{Baranov2}.
\par
The role of such surface excitations was first discussed by 
Baranov\cite{Baranov2} in connection with a BCS superfluid gas in a trap,
using the WKB semi-classical solution of the BdG equations.
More recent papers in the context of the BCS-BEC crossover 
are by Kinnunen, Rodriguez, and T\"orm\"a\cite{Torma} using LDA, 
and by Heiselberg\cite{Heiselberg} using the WKB approximation.
We note that bound state energies decrease
when the effective potential width $d$ in Fig. \ref{fig15}(b) increases. 
In a trapped Fermi gas, the analogous effective potential well becomes
wider as we enter the crossover regime because the spatial width of the
composite order
parameter ${\tilde \Delta}(r)$ 
shrinks in this region, as shown in Fig. \ref{fig7}. 
In addition, the decrease of chemical potential $\mu$ leads to
a more gradual slope of the diagonal potential 
$V_{\rm trap}^{\rm eff}(r)$ around $V_{\rm trap}^{\rm eff}(r)=0$.
These are the reasons
why $E_g$ slightly decreases (see the inset of Fig. \ref{fig11}) 
with increasing $(k_{\rm F}a_s)^{-1}$, in the
region $0.65>(k_{\rm F}a_s)^{-1}>0$. We note that the lowest Andreev bound
state energy level does not determine $E_g$ once we are in the BEC region ($\mu<0$).
In this case, $2E_g$ is dominated
by the dissociation energy of tightly bound molecules and thus 
we find $E_g\sim |\mu|\sim|\nu|$ ($\nu\to-\infty$) in Fig. \ref{fig11}.
\par
In connection with the WKB approximation\cite{Baranov2,Heiselberg}, we remark
that this should be quite good in the extreme BCS limit, where there is still a
well-defined Fermi surface. However, this approach seems less justified
in the crossover region, where the important single-particle states are
no longer close to the Fermi energy. Further studies are needed of the validity of the
semi-classical approximation to the BdG equations in the crossover
region.
\vskip3mm
\section{rf-tunneling spectroscopy in a trapped Fermi superfluid}
\par
In this section, we discuss recent work using 
rf-tunneling spectroscopy. 
As with any tunneling experiment, this clearly gives information about
the single-particle
excitations in fermion 
superfluids\cite{Chin,Torma,Torma2,Torma3,Torma4,Heiselberg,Luxat,Burnett}. 
In particular, one can extract information about both the energy gap of the
single particle excitations as well as the value of the pair 
order parameter, but this can only be done by comparison with theoretical
calculations.
The rf-tunneling spectroscopy cross-section is calculated by considering the 
tunneling current induced by
laser radiation. In the rotational wave approximation, 
the tunneling Hamiltonian is given by\cite{Torma,Torma2,Torma3,Torma4}
\begin{eqnarray}
H_t
&=&
H^F_t+H^M_t
\nonumber
\\
&=&
t_F\int d{\bf r}
\Bigl[
e^{i({\bf q}_L\cdot{\bf r}-\omega_L t)}
\Psi_a^\dagger({\bf r})
\Psi_\uparrow({\bf r})+h.c.
\Bigr]
\nonumber
\\
&+&
t_M\int d{\bf r}
\Bigl[
e^{i({\bf q}_L\cdot{\bf r}-\omega_L t)}
\Psi_a^\dagger({\bf r})
\Psi_\downarrow^\dagger({\bf r})
\Phi({\bf r})+h.c.
\Bigr].
\label{eq.5.1}
\end{eqnarray} 
Here ${\bf q}_L$ and $\omega_L$ represent the momentum and
frequency of the laser light, respectively. The first term $H_t^F$
describes the usual tunneling of an atom in the $\downarrow$-spin
state to another hyper-states described by the fermion field
operator $\Psi_a({\bf r})$, with a strength given by the tunneling matrix element
$t_F$. (We assume that $t_F$ has no spatial dependence.) This new
hyperfine state $\Psi_a$ is described by the Hamiltonian
\begin{equation}
H_a=\int d{\bf r}
\Psi_a^\dagger({\bf r})
\Bigl[
-{\nabla^2 \over 2m}+\omega_a-\mu_a+V_{\rm trap}^F
\Bigr]
\Psi_a({\bf r}),
\label{eq.5.2}
\end{equation}
where $\omega_a$ is the threshold energy and $\mu_a$ is the chemical
potential of this state. In Eq. (\ref{eq.5.2}), 
we assume that atoms in this state are non-interacting and feel
the same trap frequency as the other hyperfine states
$\uparrow,\downarrow$.
The second term $H_t^M$ in Eq. (\ref{eq.5.1}) describes the tunneling
into state $\Psi_a$ associated with the 
dissociation of a bosonic molecule, with the matrix
element $t_M$. This process is the signature of a Feshbach
resonance. In the superfluid phase at $T=0$, the Bose quantum
field operator $\Phi({\bf r})$ can be replaced by the macroscopic wavefunction
$\phi_M({\bf r})$.
In this case, the molecule component in Eq. (\ref{eq.5.1}) reduces to
\begin{eqnarray}
H_t^M=
t_M\int d{\bf r}
\Bigl[
e^{i({\bf q}_L\cdot{\bf r}-\omega_L t)}
\phi_M({\bf r})\Psi_a^\dagger({\bf r})
\Psi_\downarrow^\dagger({\bf r})+h.c.
\Bigr].
\label{eq.5.3}
\end{eqnarray} 
\par
The tunneling current operator is obtained from ${\hat I}(t)={\dot
  N}_a(t)=i[H,N_a(t)]$, where $N_a\equiv\int d{\bf
  r}\Psi^\dagger_a({\bf r})\Psi_a({\bf r})$ is the number operator of
the $\Psi_a$-state (for details, see, for example, Chap. 9 of Ref. \cite{Mahan}). 
The resulting current operators originating from
$H_t^F$ and $H_t^M$ are given by, respectively,
\begin{equation}
{\hat I}_F(t)\equiv -i
t_F\int d{\bf r}
\Bigl[
e^{i({\bf q}_L\cdot{\bf r}-\omega_L t)}
\Psi_a^\dagger({\bf r})
\Psi_\uparrow({\bf r})-h.c.
\Bigr],
\label{eq.5.4}
\end{equation}
\begin{equation}
{\hat I}_M(t)\equiv -i
t_M\int d{\bf r}
\Bigl[
e^{i({\bf q}_L\cdot{\bf r}-\omega_L t)}
\phi_M({\bf r})\Psi_a^\dagger({\bf r})
\Psi_\downarrow^\dagger({\bf r})-h.c.
\Bigr].
\label{eq.5.5}
\end{equation}
Assuming the tunneling matrix elements $t_F$ and $t_M$ are small, we can
evaluate the tunneling current using first order perturbation in 
$H_t$. The current associated with the usual
tunneling term $H_t^F$ induced by the rf-field is given by (we set ${\bf q}_L=0$)
\begin{equation}
I_F(\omega)=
\langle{\hat I}_F(\omega)\rangle=
2t_F^2
Im\int d{\bf r}d{\bf r}'
\Pi_F({\bf r},{\bf r}',-\omega),
\label{eq.5.7}
\end{equation}
where
\begin{equation}
\omega\equiv\omega_L-\omega_a-\mu+\mu_a
\label{eq.5.7b}
\end{equation}
defines the effective {\it detuning frequency}.
The two-particle Green's function $\Pi_F({\bf r},{\bf r}',\omega)$ 
in Eq. (\ref{eq.5.7}) is obtained from the
analytic continuation\cite{Mahan} of the thermal Green's function
\begin{eqnarray}
\Pi_F({\bf r},{\bf r}',i\nu_n)
&\equiv&
-\int_0^\beta d\tau
e^{i\nu_n\tau}
\langle
T_\tau
\{
\Psi_a^\dagger({\bf r},\tau)\Psi_\uparrow({\bf r},\tau)
\Psi_\uparrow^\dagger({\bf r}')\Psi_a({\bf r}')
\}
\rangle
\nonumber
\\
&=&
{1 \over \beta}\sum_{i\omega_m}
G_{11}({\bf r},{\bf r}',i\omega_m+i\nu_n)
G_a({\bf r}',{\bf r},i\omega_m).
\label{eq.5.8}
\end{eqnarray}
Here, $G_a$ is the single-particle thermal 
Green's function for the $\Psi_a$ state quantum field operator, 
\begin{equation}
G_a({\bf r},{\bf r}',i\omega_m)=
\sum_{nlm}
{Y_{lm}({\hat \theta})
R_{nl}^F(r)
R_{nl}^F(r')
Y_{lm}^*({\hat \theta}')
\over 
i\omega_m-\xi_{nl}^F}.
\label{eq.5.9}
\end{equation}
Here $\xi^F_{nl}=\omega_0(2n+l+3/2)-\mu_a$ describe the energy levels of
the $\Psi_a$ atomic hyperfine states.
We note that Eq. (\ref{eq.5.9}) does not explicitly involve the threshold energy
$\omega_a$ or the chemical potential $\mu_a$, because these have been
included in the effective detuning $\omega$ as defined in Eq. (\ref{eq.5.7b}). 
Hopefully, there will be no confusion with the label of the Fermi Matsubara frequency
$\omega_m=\pi T(2m+1)$ and the azimuthal quantum number $m$.
\par
We substitute the diagonal Green's functions $G_{11}$ in 
Eq. (\ref{eq.2.29}) and $G_{a}$ in Eq. (\ref{eq.5.9}) into
Eq. (\ref{eq.5.8}), and carry out the Matsubara $\omega_m$-frequency 
summation in the usual way\cite{Mahan}. After doing the analytic continuation to real frequencies, we obtain (at $T=0$ and $\omega>0$)
\begin{eqnarray}
I_F(\omega)=
2\pi t_F^2
\sum_l (2l+1)
\sum_{j=0}^{N_l}\sum_{n=0}^{N_l}
|W^l_{n+1,{\bar N}_l+j}|^2
\Theta(\xi^F_{nl})
\delta(\xi_{nl}^F+E_{jl}^F-\omega),
\label{eq.5.10}
\end{eqnarray}
where $E_{jl}^F$ is the energy eigenvalue 
given by the self-consistent solutions of 
the BdG equations.
\par
The tunneling current from molecules
$I_M(\omega)=\langle {\hat I}_M(\omega)\rangle$ can be calculated in
the same way. Within first order perturbation in $H_f^M$, we find
\begin{equation}
I_M(\omega)=2t_M^2
Im\int d{\bf r}d{\bf r}'\phi_m({\bf r})\phi_m({\bf r}')
\Pi_B({\bf r},{\bf r}',-\omega),
\label{eq.5.7bb}
\end{equation}
where
\begin{eqnarray}
\Pi_B({\bf r},{\bf r}',i\nu_n)
&\equiv&
-\int_0^\beta d\tau
e^{i\nu_n\tau}
\langle
T_\tau
\{
\Psi_a^\dagger({\bf r},\tau)\Psi^\dagger_\downarrow({\bf r},\tau)
\Psi_\downarrow({\bf r}')\Psi_a({\bf r}')
\}
\rangle
\nonumber
\\
&=&
{1 \over \beta}\sum_{i\omega_m}
G_{22}({\bf r},{\bf r}',i\omega_m+i\nu_n)
G_a({\bf r}',{\bf r},i\omega_m).
\label{eq.5.8b}
\end{eqnarray}
Carrying out the $\omega_m$-frequency summation in Eq. (\ref{eq.5.8b}) and the 
analytic 
continuation, we obtain
\begin{eqnarray}
I_M(\omega)=
2\pi t_M^2
\sum_l (2l+1)
\sum_{j=0}^{N_l}\sum_{n=0}^{N_l}
|\Xi^F_{nj}|^2
\Theta(\xi^F_{nl})
\delta(\xi_{nl}^F+E_{jl}^F-\omega),
\label{eq.5.11}
\end{eqnarray}
where the matrix element is given by 
\begin{equation}
\Xi^F_{nl}\equiv
\sum_{n'=0}^{N_l}
W^l_{{\bar N}_l+n',{\bar N}_l+j}
\int_0^\infty r^2dr
R_{nl}^F(r)\phi_M(r)R^F_{n'l}(r).
\label{eq.5.12}
\end{equation}
The analogue of these results were first worked out by T\"orm\"a and
co-workers\cite{Torma,Torma2,Torma3,Torma4}, both for a uniform gas and
a trapped gas within the 
LDA. Refs. \cite{Torma,Torma4} only include the molecular tunneling current
originating from the dissociation of {\it excited} molecules, 
which only becomes important at finite temperatures close to $T_{\rm c}$. 
In the case of $T=0$, which we are considering, 
all the molecules are Bose-condensed. 
In this case, the contribution in Eq (\ref{eq.5.11}) associated with the
dissociation of the {\it Bose-condensed} molecules gives the dominant
contribution to the molecular current. 
\par
In order to illustrate the physics of the preceding expressions for
$I_{\rm F}(\omega)$ and $I_M(\omega)$, it is useful to
consider a {\it uniform} superfluid 
Fermi gas at $T=0$. In this case, it is convenient
to evaluate $I_F(\omega)$ and $I_M(\omega)$ in momentum space. In
a uniform gas, the two-particle Green's function $\Pi$ in
Eq. (\ref{eq.5.7}) only depends on the relative coordinate as
$\Pi({\bf r}-{\bf r}',-\omega)$, so that Eq. (\ref{eq.5.7}) can be
written as
\begin{equation}
I_F(\omega)=2t_F^2 {1 \over \beta}\sum_{{\bf p},i\omega_m}Im
\Bigl[
G_{11}({\bf p},i\omega_m+i\nu_n)G_a({\bf p},i\omega_m)
\Bigr],
\label{eq.5.12b}
\end{equation}
where $G_{11}$ is given in Eq. (\ref{eq.10}), and $G_a^{-1}({\bf
  p},i\omega_m)\equiv i\omega_m-\xi_{\bf p}$ is the single-particle
Green's function of a free uniform Fermi gas of atoms in the
hyperfine state `a'. After doing the
$\omega_m$- and ${\bf p}$-summations, one finds\cite{Torma2}
\begin{equation}
I_F(\omega)=\pi t_F^2
\rho
\Bigl(
\Omega=
{1 \over 2}{\omega^2-{\tilde \Delta}^2 \over \omega}+\mu
\Bigr)
{{\tilde \Delta}^2 \over \omega^2}
\Theta(\omega-{\tilde \Delta})
\Theta
\Bigl(
{1 \over 2}{\omega^2-{\tilde \Delta}^2 \over \omega}+\mu
\Bigr).
\label{eq.5.13}
\end{equation}
Here, $\rho(\Omega)\propto\sqrt{\Omega}$ 
is normal state density-of-states given below Eq. (\ref{eq.y1}).
Equation (\ref{eq.5.13}) clearly has a peak at the 
energy gap given by $\omega=|{\tilde
  \Delta}|$ as long as $\mu>0$ (BCS region). On the other hand, 
when $\mu<0$ (BEC region), the
threshold energy is given by the last factor in Eq. (\ref{eq.5.13}),
\begin{equation}
\omega_{\rm th}=|\mu|+\sqrt{\mu^2+{\tilde \Delta}^2},
\label{eq.5.14}
\end{equation}
rather than $\omega_{\rm th}=|{\tilde \Delta}|$.
In this BEC limit, the expression in Eq. (\ref{eq.5.14}) approaches
the binding energy of a molecular boson as
$\omega_{\rm th}\to2|\mu|\simeq2|\nu|$, which is 
twice as large as the threshold energy (or energy gap) of the
single-particle Fermi excitations (in a uniform gas). 
As one might expect, the threshold energy of the
rf-induced tunneling current $I_{\rm F}(\omega)$ continuously changes from 
the single-particle excitation gap $|{\tilde \Delta|}$ in
the BCS regime to the threshold of the two-particle continuum in the BEC
limit. In a trap, the tunneling current 
$I_F(\omega)$ in
LDA\cite{Torma} is given simply by the 
spatial integration over Eq. (\ref{eq.5.13}), where
${\tilde \Delta}({\bf r})$ and $\mu({\bf r})\equiv\mu-{1 \over 2}m\omega_0^2r^2$ 
now depend on the position ${\bf r}$. The sharp
peak at the excitation threshold then becomes broadened, as first discussed in
Ref.\cite{Torma}.
\par
Figure \ref{fig16} shows our calculated results for the rf-induced current 
$I_F(\omega)$ in the BCS-BEC crossover at
$T=0$ in a {\it trapped} gas. Since we take into account the discrete energy
levels of the harmonic trap potential (see Fig. \ref{fig10}), 
the spectrum shows rapid oscillations. 
In panel (a) describing the BCS region, 
one sees that the lowest tunneling current frequency is at 
$\omega\simeq 0.08\varepsilon_{\rm F}$\cite{note0704}. This
corresponds to the single-particle excitation gap
$E_g$ discussed in Section VI. 
A broad peak is also evident, centered at $\omega\sim 0.3\varepsilon_F$. This peak
energy is seen to decrease as one enters the crossover region 
[panel (b) of Fig. \ref{fig16}]. 
Since the profile of the composite order parameter
shrinks 
and the chemical potential $\mu$ decreases 
in this regime (see Figs. \ref{fig6} and \ref{fig7}), 
the width of the combined
potential well $[{\tilde \Delta}(r)+V_{\rm eff}^{\rm trap}(r)$, 
see also the inset in Fig. \ref{fig14}(b)] increases. 
In this case, a large number of low-energy 
excitations appear, which are localized at the minimum of 
the potential well. These low energy excitations lead to 
the large value of $I_{\rm F}(\omega)$ at low $\omega$, as shown in 
Fig. \ref{fig16}(b). In contrast, once we enter the BEC regime, the
single-particle
excitation gap $E_g$ quickly becomes large (see Fig. \ref{fig11}). This shows up in
the rf-tunneling currents in panels (c) and (d), where the
low-energy Andreev surface excitations no
longer determine the pair threshold in $I_{\rm F}(\omega)$, even though 
these states still exist. 
In Fig. \ref{fig16}(d), we see the threshold energy of the
Fermi spectrum becomes quite large, at $\omega_{\rm th}\simeq 2\varepsilon_{\rm F}$,
a result of being in the BEC region. 
This value is approximately twice as large as $E_g$ (see Fig. \ref{fig11}) for 
this value of $\nu$, and is consistent with the discussion given above
for the case of a uniform superfluid.
\par
For comparison, Fig. \ref{fig17} shows the 
rf-tunneling spectroscopy calculated using the LDA, which is 
given by [see Eq. (\ref{eq.5.13})]
\begin{equation}
I_F(\omega)=\pi t_F^2
\int d{\bf r}
\rho
\Bigl(\Omega=
{1 \over 2}{\omega^2-{\tilde \Delta}(r)^2 \over \omega}+\mu(r)
\Bigr)
{{\tilde \Delta}(r)^2 \over \omega^2}
\Theta(\omega-{\tilde \Delta}(r))
\Theta
\Bigl(
{1 \over 2}{\omega^2-{\tilde \Delta}(r)^2 \over \omega}+\mu(r)
\Bigr).
\label{eq.5.13z}
\end{equation}
Here $\mu(r)=\mu-{1 \over 2}m\omega_0^2r^2+{U \over 2}n_{\rm F}(r)$, where ${U \over 2}n_{\rm F}(r)$ is the Hartree term. The values of $\mu$, ${\tilde \Delta}(r)$, and $n_{\rm F}(r)$ which are used in Eq. (\ref{eq.5.13z}) 
are obtained by solving the BdG coupled equations
in a self-consistent manner, as discussed in earlier sections.
\par
Comparing Fig. \ref{fig17} with Fig. \ref{fig16}, we find that 
when these self-consistent solutions of the BdG equations are used, 
the LDA gives a good overall approximation 
in the whole crossover regime for the rf-tunneling spectroscopy.
However, it does not give 
the {\it true} energy gap ($E_g\sim 0.08\varepsilon_{\rm F}\ll\varepsilon_{\rm F}$) 
in the BCS regime, in contrast to the microscopic results shown in 
Figs. \ref{fig16}(a) and (b).
Since this energy gap $E_g$ 
originates from the Andreev bound states, 
it is no surprise that an LDA calculation does not reproduce it.
\par
Although the peak energy in the tunneling current is very small in Fig. \ref{fig16}(b), this
does {\it not} mean that 
the ``average'' magnitude of the composite order
parameter is small. When we extract the contribution coming from 
the central region of the trap ($0\le r\le r_c$, where $r_c\ll R_{\rm F}$),
$I_{\rm F}(\omega)$ has a peak at a high energy, reflecting the large magnitude
of ${\tilde \Delta}(r\sim 0)$. This is shown in Fig. \ref{fig17}(b) for the case
$r_c=0.3R_{\rm F}$ (line 1), where $I_{\rm F}^{LDA}(\omega)$ has a peak around
$\omega\sim\varepsilon_{\rm F}$. In this restricted spatial 
region ($0\le r\le 0.3R_{\rm F}$), 
${\tilde \Delta}(r)$ is of the order of the Fermi energy, as shown 
in the inset of Fig. \ref{fig17}(b). As one increases $r_c$,
the contribution of 
the low-energy excitations localized around the surface region of the cloud 
begins to ``hide" this high energy peak. When we take $r_c=0.6R_{\rm F}$ in Fig. \ref{fig17}(b)
(the line 2), the peak in the spectrum is still dominated by the 
low energy excitations. 
It would be very useful in future experiments
if one could measure the rf-tunneling current from the central region of the trap.
In principle, such selective measurements could give detailed information about
the spatial dependence of ${\tilde \Delta}(r)$.
\par
When the profile of the composite order parameter spreads out up
to $R_{\rm F}$ (namely, when ${\tilde \Delta}(r)$ is large even close to
the trap edge), one expects that the effect of high energy excitations with an
energy gap close to ${\tilde \Delta}(r=0)$ will dominate over 
low-energy excitations in the rf-spectrum $I_{\rm F}(\omega)$.
To confirm this expectation, we show a calculation in Fig. \ref{fig18}
using an ad-hoc model for ${\tilde \Delta}(r)$ shown in the inset.
One finds that, as expected, 
the rf-tunneling spectrum for this broad order parameter has a high-energy peak at 
$\omega\sim{\tilde \Delta}(0)\sim\varepsilon_{\rm F}$. 
This kind of slowly decreasing ${\tilde \Delta}(r)$ might be obtained
when the effective repulsive molecule-molecule is strong, which our calculations 
have ignored. We also note that 
our explicit calculations are for a {\it narrow} Feshbach resonance, in which case
molecules are dominant in the crossover region. 
[Our self-consistent expression for ${\tilde \Delta}(r)$ is shown in 
the inset of Fig. \ref{fig17}(b).] 
In a {\it broad} Feshbach resonance, where Cooper-pairs
are dominant in the crossover regime and the size of these Cooper-pairs is still
fairly large, such a broad profile of ${\tilde \Delta}(r)$ may be possible.
In the recent calculations of the rf-tunneling by Kinnunen et al.\cite{Torma},
a high-energy peak was obtained in the low temperature limit. The difference
between the result obtained in Ref. \cite{Torma} 
and ours (Figs. \ref{fig16} and \ref{fig17}) thus seems largely due to
the difference in the spatial 
profile of the composite order parameter ${\tilde \Delta}(r)$
used in the two calculations.
\par
As shown in Fig. \ref{fig18} and the discussion above, 
the rf-tunneling current $I_F(\omega)$ is very
sensitive to the detailed spatial structure of ${\tilde \Delta}(r)$. This is simply
because the factor $\omega^{-2}$ in $I_F(\omega)$ [see
Eq. (\ref{eq.5.13})] tends to emphasize the role of the low energy
excitations of the superfluid gas. 
To extract information about the high-energy region and 
the magnitude of ${\tilde \Delta}(0)$ in the center of the trap,
it is useful to consider the function ${\tilde
  I}_F(\omega)\equiv\omega^2 I_{\rm F}(\omega)$. In a uniform gas, 
when we neglect the energy dependence of the normal state density-of-states 
$\rho(\omega)$ for simplicity, we find that 
${\tilde I}_F(\omega)\propto \Theta(\omega-|{\tilde \Delta}|)$, so that
one can directly
determine the magnitude of ${\tilde \Delta}$ from the energy at which
the spectrum shows a sudden jump. 
Using the LDA, this discontinuity
at $\omega=|{\tilde \Delta}|$ is broadened in a trap due to the
  inhomogeneity of ${\tilde \Delta}(r)$. However, one can still expect that 
${\tilde I}_F(\omega)$ would start to decrease from 
an energy of the order of the maximum gap, since most atoms are at the 
center of the trap. 
This behavior is clearly shown in Fig. \ref{fig19}, where ${\tilde I}_F(\omega)$
is seen to be suppressed for $\omega\lesssim{\tilde \Delta}(r=0)$.
At the center of the trap, ${\tilde \Delta}(0)$ is of the order 
$1.5\sim 2\varepsilon_{\rm F}$ (see Fig. \ref{fig13}). We conclude that a plot of 
${\tilde I}_F(\omega)$ can be used to estimate the magnitude of the
composite order parameter at the center of the trap, even if there is a low-energy
peak in $I_{\rm F}(\omega)$.
\par
Figure \ref{fig20} shows the spectrum at $T=0$ of the molecular current $I_M(\omega)$
given by Eq. (\ref{eq.5.11}). 
This Bose rf-tunneling spectrum is seen to be very similar to 
${\tilde I}_F(\omega)$, as defined above (see Fig. \ref{fig19}). 
In particular, we see that the frequency where
the rf-spectrum starts to be suppressed corresponds quite closely
to the maximum value of the order parameter
${\tilde \Delta}(r=0)$ 
at the center of the trap (as shown by the arrows in Fig. \ref{fig20}). In a
{\it uniform} gas, one can calculate $I_{\rm M}(\omega)$ explicitly,
\begin{equation}
I_M(\omega)=\pi t_M^2\phi_M^2
\rho
\Bigl(\Omega=
{1 \over 2}{\omega^2-{\tilde \Delta}^2 \over \omega}+\mu
\Bigr)
\Theta(\omega-{\tilde \Delta})
\Theta
\Bigl(
{1 \over 2}{\omega^2-{\tilde \Delta}^2 \over \omega}+\mu
\Bigr).
\label{eq.5.15}
\end{equation}
Thus in a uniform gas, we find that $I_M(\omega)\propto{\tilde I}_F(\omega)$,
since the factor $\omega^{-2}$ in Eq. (\ref{eq.5.13}) 
is not present in $I_M(\omega)$.
Thus, measuring the molecular dissociation current $I_M(\omega)$ 
would appear to give a more direct way of probing the spectrum of the
high energy excitations, and hence
the magnitude of the composite order parameter ${\tilde \Delta}(r=0)$. 
In recent
discussions\cite{Torma4}, a general expression for 
the molecular tunneling current was discussed, 
but it was not evaluated in Ref.\cite{Torma}. This neglect was
justified by the assumption that in the BCS region of interest,
the number of molecules was small. 
For a broad Feshbach resonance considered in Ref. \cite{Torma}, this may be
correct but further studies are needed.
\par
Since we can eliminate the effect of low-energy excitations on the rf-spectroscopy
and extract the information about ${\tilde \Delta}(0)$ by considering 
${\tilde I}_F(\omega)$ and $I_M(\omega)$, it is interesting to see where
this ``hidden peak" {\it coming from this high-energy contribution} 
is in the spectrum of
$I_{\rm F}(\omega)$
in cases where it may be 
masked by the low energy spectral weight (as in Fig. \ref{fig16}).
For this purpose, we can simply model the ${\tilde I}_F(\omega)$ 
spectrum using the Lorentzian form
\begin{equation}
{\tilde I}_F(\omega)={C\omega^2 \over (\omega-\omega_p)^2+\Gamma^2}.
\label{rf1}
\end{equation}
The parameters $C$, $\Gamma$, and $\omega_p$ 
can then be determined from the best fit to the calculated ${\tilde I}_F(\omega)$
{\it around the high-energy region} $\omega\sim {\tilde \Delta}(0)$. 
As an example, such a fit is shown by the dashed line in Fig. \ref{fig19}.
The corresponding rf-tunneling spectrum $I_{\rm F}(\omega)$ 
is shown by the dashed line in Fig. \ref{fig16}(b). 
The energy of the ``hidden" broad peak in $I_{\rm F}(\omega)$ 
is given by the value of $\omega_p$
found by this procedure. 
The results for $\omega_p$ using this procedure 
are plotted in Fig. \ref{fig21}.
We find reasonable agreement with the experimental data for $^6$Li\cite{Chin},
especially for the case ${\bar U}=0.35\varepsilon_{\rm F}$. 
Our peak energy at $(k_{\rm F}a_s)^{-1}=0$ occurs at 
$\omega\simeq0.3\varepsilon_{\rm F}$, which 
is in agreement with the results of the recent 
theoretical analysis using the LDA by T\"orm\"a and 
co-workers\cite{Torma}, who found a broad peak at $\omega\sim0.3\varepsilon_{\rm F}$
near the unitarity limit.
We also recall that the calculation in Ref. \cite{Torma}
were for a broad resonance. It would be very useful to have the calculations
in Ref. \cite{Torma} extended to cover the whole crossover region (results were only
reported for $\nu=0.5\varepsilon_{\rm F}$).
\par
In the recent rf-spectroscopy data at finite temperatures of Grimm and
co-workers\cite{Chin}, one finds a strong narrow peak at zero detuning 
as well as
a broad peak at positive detuning. As the temperature decreases, the
spectral weight shifts to the broad peak. It is argued in Ref. \cite{Chin}
that the peak at zero
detuning
is due to unpaired or free Fermi atoms at the edge of the trap, which
have no energy gap. The fact that the peak is narrow is further argued to be
evidence that these states come from the region of low density, 
consistent with
negligible mean-field broadening. The pioneering theoretical work of T\"orm\"a
and
co-workers based on the LDA\cite{Torma} 
leads to the same interpretation. 
However, as discussed at the end of Sec. V, the LDA at finite
temperatures incorrectly predicts a region at the edge of the trap
where the order parameter ${\tilde
  \Delta}(r)$ has vanished even though the density of atoms $n(r)$ is
still finite (see Fig. 1 of Ref. \cite{Torma}). 
In fact, the BdG equations show that the {\it entire} 
trapped gas is in a superfluid state below $T_{\rm c}$,
with ${\tilde \Delta}(r)$ being finite everywhere where $n(r)$
is finite.
\par
As we discussed in Section VI, very low energy states (with a finite but small
excitation gap $E_g$) arise which are 
localized in the low density tail of the superfluid gas,
the analogue of Andreev states\cite{dG2,Baranov2}. 
These in-gap surface states\cite{Torma3,Heiselberg} 
are a true signature of the Fermi superfluid phase, 
since they involve a coherent mixture of 
particle and hole components\cite{dG2}. 
In particular, these excitations exist even at $T=0$. 
At finite temperatures below $T_{\rm c}$, the rf-tunneling
current $I_{\rm F}(\omega)$ involves two 
contributions, coming from thermally excited
Bogoliubov quasi-particles and from the Cooper-pair condensate. The former contribution
disappears at $T=0$, because all the atoms are paired and Bose-condensed,
and there are no excitations. 
However, the latter contribution exists even at $T=0$. 
\par
The low-energy
rf-current spectrum $I_{\rm F}(\omega)$ is in fact dominated by 
excitations from the condensate associated with 
 the low-energy Andreev or in-gap states.
Indeed, we can clearly see the {\it true} very small excitation gap 
$E_g\ll\varepsilon_{\rm F}$ from the lowest energy peaks 
in panels (a) and (b) of Fig. \ref{fig16}.
T\"orm\"a and co-workers\cite{Torma} have argued that the 
normal phase atoms at the trap edge in their LDA calculation can be
viewed as a crude approximation to these in-gap low energy states, which arise in
a more accurate theory based on the BdG equations.
However, in an LDA analysis, the low frequency (free atom)
peak is predicted to disappear at very low temperatures when
the entire trapped gas becomes superfluid. This picture is different from
what our microscopic calculations give, as discussed above. 
For the same reason, we feel that the data in Ref. \cite{Chin} does not
give any convincing 
evidence for the pseudogap phase discussed in Refs. \cite{Stajic,Levin}.
\par
A proper discussion of the low energy states 
in the edge of a trapped superfluid gas requires the kind of
microscopic calculations presented in this paper.
One is losing a huge amount of physics by thinking of the
central peak (at zero detuning) simply as the contribution from unpaired atoms
(or atoms feeling the effect of a ``pseudogap''). This low energy peak at 
$\omega\ll\varepsilon_{\rm F}$ is due to the characteristic
low energy states ($\sim E_g\ll\varepsilon_{\rm F}$)
of a trapped superfluid gas in the BCS region (see Fig. \ref{fig10}). 
The BEC region is quite different. We
recall from Fig. \ref{fig11} that $E_g$ rapidly increases as we pass from the
BCS to the BEC regions. In the BEC region, the large energy gap $E_g$ is 
determined by the magnitude of the chemical potential, rather than the 
low energy Andreev in-gap states. It would be very useful to attempt a higher 
resolution study of the ``unshifted peak" 
in the rf-induced current measurements as presented in 
Ref. \cite{Chin}.
This could give more detailed information about the low-energy in-gap 
excitations of a 
trapped Fermi superfluid in the BCS region.
\par
\vskip5mm
\section{Concluding Remarks}
\vskip2mm
In this paper, we have presented  a detailed study of 
the equilibrium properties and single-particle 
excitations in the BCS-BEC crossover regime of a trapped Fermi gas
with a Feshbach resonance, at $T=0$. 
We extended the crossover theory developed
by Leggett\cite{Leggett} to include the effect of a Feshbach resonance in a trapped
Fermi gas. In our work, the composite order parameter ${\tilde \Delta}(r)$, 
the atomic density
profile $n_{\rm F}(r)$, and the 
chemical potential $\mu$ are all calculated self-consistently. 
Our theory does not use the LDA, but works with the correct eigenstates of the
harmonic trapping potential given by the Bogoliubov-de Gennes 
(BdG) coupled equations. 
In a uniform BCS Fermi superfluid, the single-particle excitations have an energy gap
which is equal to the pair potential $\Delta$. In contrast, in the BCS-BEC
crossover region, this single-particle energy gap is not directly
proportional to the pair potential ${\tilde \Delta}$. 
In a trapped Fermi superfluid, there is never a simple relation between the
energy gap $E_g$ and the underlying spatially varying 
 order parameter ${\tilde \Delta}(r)$. One of the themes of our paper
is how to extract information about both these quantities.
\par
We showed that spatially-dependent local 
density and the order
parameter become more localized at the center of the trap 
as one decreases the threshold energy ($2\nu$) of the Feshbach
resonance. This reflects the fact that the character of the particles is continuously 
changing, from unpaired Fermi atoms to bound states (molecules) 
associated with the Feshbach
resonance. The threshold energy $E_{\rm g}$ of single-particle excitations
was shown to be much smaller than the magnitude of composite order
parameter at the center of the trap 
in the crossover region, where ${\tilde \Delta}(r=0)\sim \varepsilon_{\rm F}$. 
This is because $E_{\rm g}$ in this region 
is determined by the lowest Andreev (or in-gap) bound states\cite{Baranov2} 
near the bottom of the combined potential 
well composed of the off-diagonal pair potential and the trap potential. 
We have emphasized that these states are a characteristic signature
of a trapped superfluid Fermi gas and hence are of special interest.
\par
We also used our results for the single-particle excitation spectrum to discuss 
recent rf-tunneling experiments. 
As discussed recently\cite{Chin,Torma}, the data at finite temperatures in the crossover region can be usefully described in terms of a narrow unshifted peak ($\omega\sim 0$)
and a broad peak at a detuning frequency $\omega$ comparable to the expected
pair potential ${\tilde \Delta}(r=0)$ at the center of the trap.
While LDA calculations for the case $\nu=0.5\varepsilon_{\rm F}$\cite{Torma}
appear to confirm this kind of rf-spectrum, our present calculations
based on the explicit solutions of the BdG equations at $T=0$ lead to
somewhat different predictions about the low $\omega$ region.
We have verified that our results are essentially reproduced by an LDA calculation
if we base it our self-consistent values of $n_{\rm F}(r)$ and 
${\tilde \Delta}(r)$ given by the BdG equations. 
The major difference between our results and those in 
Refs. \cite{Chin,Torma} is that we find a strong low frequency contribution to 
the fermionic tunneling current in the BCS-BEC crossover region.
However, we find that the rf-tunneling current $I_{\rm F}(\omega)$ 
is very dependent on the
precise spatial dependence of ${\tilde \Delta}(r)$.
This dependence is good news since it means that fits to the rf-tunneling data
may be used in the future 
to obtain detailed information about the spatial dependence
of the composite order parameter ${\tilde \Delta}(r)$ in the crossover region.
\par
Further work is needed to clarify the role of the Andreev bound states in 
determining the low frequency peak in the rf-spectrum data.
These single-particle states continue to exist even at $T=0$, and
are clearly left out of an LDA type calculation, which is based on the results for a 
uniform Fermi superfluid.
\par
In Sec. VII, we showed that the low frequency peak in the 
rf-spectrum was due to contribution from the edge of the trap, where
${\tilde \Delta}(r)$ is very small [see Fig. \ref{fig17}(b)].
Subtracting out this ``surface" contribution, the remaining rf-spectrum was peaked at
high energies comparable to the value of ${\tilde \Delta}(r=0)$ at the
center of the trap. These peak energies were in reasonable agreement
with the observed position of the broad peak in the tunneling data reported in
Ref. \cite{Chin}, as shown in Fig. \ref{fig21}.
\par
We have also evaluated the rf-spectrum due to the
current produced by dissociation of molecules (see Fig. \ref{fig20})
and found that it did not have an unshifted component at $\omega\sim0$. 
Thus it gives a more
direct measurement of ${\tilde \Delta}(r=0)$ at the center of the trap.
This contribution, which was discussed but 
not explicitly evaluated in Ref. \cite{Torma}, deserves
further study, both experimentally and theoretically.
In future work, we hope to extend our present calculations to finite temperatures.
\par
Besides the single-particle properties, collective excitations in trapped Fermi 
superfluids\cite{Kinast,Bar2,Heisel2} are
also of great interest in the crossover region. 
The single-particle Green's function evaluated in the 
present paper form the basis for the calculation of
the collisionless 
collective modes in a trapped Fermi gas using linear response
theory. In a future paper\cite{Ohashi7}, 
we will extend the approach given in Ref. \cite{Ohashi4} 
for a uniform gas and discuss the quadrupole and monopole modes 
in the BCS-BEC 
crossover region\cite{Ohashi-web}. 
A detailed discussion of the Kohn mode has been recently
given in Ref.\cite{Ohashiz}.
\par
As discussed in Sec. VII, rf-tunneling spectroscopy experiments can 
give information about 
the true quasi-particle excitation gap $E_g$. However, 
very high resolution would be necessary in the BCS and crossover
regime because $E_g\sim\omega_0\ll\varepsilon_{\rm F}$
(see Figs. \ref{fig11} and \ref{fig16}). As a result, an interesting
problem remains as to 
how to measure $E_g$ in this region. An alternative method might be 
through the study of the
collective mode frequencies,
which typically have a very low energy comparable to the trap
frequency $\omega_0$ and are, in fact, bounded by the two-particle continuum at 
$2E_g$.
We show a plot of the calculated frequency of
the monopole mode in Fig. \ref{fig22} (the details are discussed in
Refs. \cite{Ohashi-web,Ohashi7}). The monopole mode frequency is seen to be
suppressed, so that it always 
lies below the two-particle continuum at $2E_g$. In particular, 
the suppression is quite striking around the 
region where $E_g$ shows a minimum as a function of $\nu$. 
This effect of the two-particle continuum on the
monopole mode frequency appears to be an attractive way of obtaining information
about the single-particle excitation gap $E_g$ in a trapped superfluid 
Fermi gas in the crossover region.
\vskip2mm
\acknowledgments
Y. O. was financially supported by a Grant-in-Aid for Scientific
research from the Ministry of Education, Culture, Sports, Science
and Technology of Japan, as well as by a University of Tsukuba Research
Project. A. G. acknowledges a research grant from NSERC of Canada.
\par 
%
%
\newpage


%
\newpage
\centerline{ }
\begin{figure}
\caption{
\label{fig1} 
(a) Superfluid transition temperature $T_{\rm c}$ as a function of the
scattering length $(k_Fa_s)^{-1}$ for a {\it uniform} gas. 
The solid line shows
the case of a {\it narrow} Feshbach resonance 
($g_{\rm r}\sqrt{n}\lesssim\varepsilon_{\rm F}$), 
while the dashed line is for the
case of a {\it broad} Feshbach resonance ($g_{\rm r}\sqrt{n}\gg\varepsilon_{\rm
  F}$), where $n$ is the density of atoms.
The peak structure around $(k_{\rm F}a_s)^{-1}\sim 0.25$ is an artifact
of the approximation and is not intrinsic\cite{Haussmann}.
The solid circle shows $T_{\rm c}$ as a function of the {\it
  two-particle} scattering length $a_s^{2b}$ in the case of 
$g_{\rm r}\sqrt{n}=20\varepsilon_{\rm F}$ (broad Feshbach resonance).
We take the cutoff $\omega_c=2\varepsilon_{\rm F}$.
(b) The number of Fermi atoms $N_{\rm F}$
and the number of Bose molecules $N_{\rm M}$ at $T_{\rm c}$ in the
BCS-BEC crossover. 
($N_M$ includes both 
Cooper-pairs and real two-body molecules.) The total number of particles is given by $N=N_{\rm F}+2N_{\rm M}$. 
Irrespective of whether one is dealing with a broad or narrow Feshbach resonance,
one notes that bound states ($N_M$) become 
dominant when $(k_{\rm F}a_s)^{-1}\gesim 0$.}
\end{figure}
\begin{figure}
\caption{
\label{fig2} 
(a) Chemical potential in the
BCS-BEC crossover for a {\it uniform} gas at $T_{\rm c}$, based on 
Ref. \cite{Ohashi1}. 
The parameters
are the same as in Fig. \ref{fig1}. Panels (b) and (c) show detailed 
character of particles at $T_{\rm c}$ for a uniform gas
in the cases of a narrow Feshbach resonance
and broad Feshbach resonance, respectively. $N_{\rm B}^{\gamma>0}$ is the number of
Feshbach molecules with a finite lifetime, and $N_{\rm B}^{\gamma=0}$ is
the number of stable Feshbach molecules.
The number of stable Cooper-pairs is $N_{\rm C}^{\gamma=0}$;
$N_{\rm C}^{\rm sc}$ is the contribution from the scattering states or Cooper-pairs with a finite lifetime. For precise definitions of $N_{\rm B}$ and $N_{\rm C}$, see Ref. \cite{Ohashi1}. 
}
\end{figure}
\begin{figure}
\caption{
\label{fig3} 
(a) Superfluid transition temperature $T_{\rm c}$ as a function of the
scattering length $(k_Fa_s)^{-1}$ for a {\it trapped} gas. The cases 
of a narrow Feshbach resonance and broad Feshbach resonance are
respectively shown by the solid line and dashed line, respectively.
In this figure, $n\equiv N/R_{\rm F}^3$, and we take the 
cutoff $\omega_c=2\varepsilon_{\rm F}$.
(b) The number of Fermi atoms $N_{\rm F}$
and the number of Bose molecules $N_{\rm M}$ at $T_{\rm c}$ in a harmonic
trap. 
}
\end{figure}
%
\begin{figure}
\caption{
\label{fig4} 
(a) Comparison between the bare attractive 
interaction $-U_{\rm eff}$ between Fermi atoms in the gap equation
[as defined in Eq. (\ref{eq.2.5d})]
and the interaction $4\pi a_s/m$ defined in Eq. (\ref{eq.bcs02}).
We take the cutoff frequency $\omega_c=161.5\omega_0$ and the Fermi energy
$\varepsilon_{\rm F}=31.5\omega_0$ in this and later figures.
(b) Relation between the scattering length $a_s$ and the bare 
threshold energy $2\nu$,
for the parameters we use in this paper. 
The Feshbach coupling strength 
is taken as ${\bar g}_{\rm r}=0.2\varepsilon_{\rm F}$.
}
\end{figure}
\begin{figure}
\caption{
\label{fig5} 
Density profile in the BCS-BEC crossover. $n_{\rm F}(r)$ 
and $n_{\rm M}(r)$
represent the Fermi atom density and Bose molecule density,
respectively. $n(r)=n_{\rm F}(r)+2n_{\rm M}(r)$ is the total
density profile. $r$ is normalized by Thomas-Fermi radius $R_{\rm
  F}=\sqrt{2\varepsilon_{\rm F}/m\omega_0^2}$ for a free Fermi gas in
a trap. We take ${\bar g}_{\rm r}=0.2\varepsilon_{\rm F}$, ${\bar
  U}=0.35\varepsilon_{\rm F}$ and $\omega_{0B}=\omega_0$. These values are
are also used in Figs. \ref{fig6}, \ref{fig7}, and \ref{fig8}. 
}
\end{figure}
\begin{figure}
\caption{
\label{fig6} 
Fermi chemical potential $\mu$ in the BCS-BEC crossover at $T=0$. 
The inset shows $\mu$ as a function of the
threshold energy $2\nu$. 
The dashed line in the inset is the lowest energy of molecular excitation
spectrum in a trap. 
}
\end{figure}
\begin{figure}
\caption{
\label{fig7} 
Spatially-dependent composite order parameter ${\tilde \Delta}(r)$
in a trap in the BCS-BEC crossover. The Cooper-pair order 
parameter $\Delta(r)$
and molecular condensate $\phi_m(r)$ are also shown. 
}
\end{figure}
\begin{figure}
\caption{
\label{fig8} 
Expansion coefficient $\alpha_n$ of the molecular condensate wavefunction
$\phi_M(r)={1 \over \sqrt{4\pi}}\sum_n\alpha_nR_{n0}^M(r)$
given by Eqs. (\ref{eq.2.17}) and (\ref{eq.2.18}). 
BEC in a non-interacting Bose gas corresponds to all $\alpha_n=0$ for $n\ne 0$. 
}
\end{figure}
\begin{figure}
\caption{
\label{fig9} 
Comparison of the calculated atomic density profile $n(r)$ 
and composite order parameter ${\tilde \Delta}(r)$. 
The results are normalized by the values at the center of
the trap. 
(a) BCS regime, and (b) BEC regime. The inset compares $n(r)$ with
${\tilde \Delta}(r)^2$ in the BEC regime.
}
\end{figure}
\begin{figure}
\caption{
\label{fig10}
Single-particle density of states (DOS) at $T=0$. 
We take 
${\bar g}_{\rm r}=0.2\varepsilon_{\rm F}$ and 
${\bar U}=0.52\varepsilon_{\rm F}$. In calculating the DOS, 
we have introduced a small imaginary part ($\Gamma=0.005\omega_0$)
in the eigenenergies. The fine structure in DOS 
is due to the discrete levels
of harmonic trapping potential.
The peaks are the single-particle Bogoliubov quasi-particle 
energies in a trapped gas.
Note that $\mu<0$ in panel (c).
}
\end{figure}
\begin{figure}
\caption{
\label{fig11}
Single-particle excitation gap $E_g$ appearing in the 
density of states $N(\omega)$ in the BCS-BEC crossover
in a trap. The dashed line shows $|\mu|$ in the negative $\mu$ region
[$(k_{\rm F}a_s)^{-1}>0.65$]. 
The region around the minimum of $E_{\rm g}$ shown in the
inset is discussed in the text. 
We take ${\bar g}_{\rm r}=0.2\varepsilon_{\rm F}$ and 
${\bar U}=0.52\varepsilon_{\rm F}$. 
}
\end{figure}
\begin{figure}
\caption{
\label{fig12}
A plot of the calculated number of atoms ($N_{\rm F}$) and molecules ($N_{\rm M}$)
at $T=0$.
The total number of atoms is given by $N=N_{\rm F}+2N_{\rm M}$.
 We find $N_{\rm M}\simeq N/2=10912/2$ in the negative $\mu$ 
region (see also Fig. \ref{fig6}).
We take ${\bar g}_{\rm r}=0.2\varepsilon_{\rm F}$ and 
${\bar U}=0.52\varepsilon_{\rm F}$. 
}
\end{figure}
\begin{figure}
\caption{
\label{fig13}
Change in the maximum value of composite order 
parameter ${\tilde \Delta}(0)$ at the
center of the trap in the crossover region. 
}
\end{figure}
\begin{figure}
\caption{
\label{fig14}
Localized density of states $N(\omega,r)$ given by Eq. (\ref{eq.2.33}) in the
crossover regime. 
The peaks correspond to the Andreev or in-gap states.
The inset in panel (b) shows the combined potential well
consisting of the composite ``off-diagonal" 
pair potential ${\tilde \Delta}(r)$ and ``diagonal" trap potential
measured from the chemical potential, given by
$V_{\rm eff}^{\rm trap}(r)\equiv(V_{\rm trap}^F(r)-\mu)\Theta(V_{\rm trap}^F(r)-\mu)$. 
Actually, atoms also feel the Hartree potential
$-{U \over 2}n_{\rm F}(r)$, but this is not plotted in the inset. 
}
\end{figure}
\begin{figure}
\caption{
\label{fig15}
(a) Combined potential well formed by the off-diagonal pair-potential ${\tilde
  \Delta}(r)$ and the diagonal trap potential $V_{\rm trap}^F(r)$.
The combined potential is given by the sum of ${\tilde \Delta}(r)$ and
$V_{\rm eff}^{\rm trap}(r)\equiv (V_{\rm trap}^F(r)-\mu)\Theta(V_{\rm trap}^F-\mu)$.
The low-energy states appear around the bottom of this effective
potential 
well. (b) Simplified
model of the combined potential well which gives rise to Andreev bound
states\cite{dG2,Baranov2}.
}
\end{figure}
\begin{figure}
\caption{
\label{fig16}
The rf-induced current $I_F(\omega)$ in the BCS-BEC crossover region
at $T=0$.
We take ${\bar g}_{\rm r}=0.2\varepsilon_{\rm F}$ and 
${\bar U}=0.52\varepsilon_{\rm F}$, and introduce a
finite imaginary part $\Gamma=0.5\omega_0$ to the 
Bogoliubov eigenstate energies to smooth out the results.
The small but finite intensity at $\omega=0$
in panels (a) and (b) is due to this imaginary part.
The dashed line in panel (b) is obtained by fitting Eq. (\ref{rf1})
to ${\tilde I}_{\rm F}(\omega)=\omega^2 I_{\rm F}(\omega)$ around 
$\omega\simeq {\tilde \Delta}(r=0)$ (see Fig. \ref{fig19}) 
to extract the high-energy ``hidden" peak associated with the
large composite order parameter ${\tilde \Delta}(r=0)$ at the
center of the trap. 
The rapid oscillations in the tunneling spectrum, 
also seen in Figs. \ref{fig19} and \ref{fig20},
originates from discrete quasiparticle energy levels
in a harmonic trap potential.
}
\end{figure}

\begin{figure}
\caption{
\label{fig17}
The rf-induced current $I_F(\omega)$ 
evaluated using the LDA. The parameters are the same as Fig. \ref{fig16}.
In calculating the spectrum, we have used the correct values of 
$n_{\rm F}(r)$, ${\tilde \Delta}(r)$, and $\mu$ obtained from the
solutions of the
BdG coupled equations. In panel (b), (1) shows the spectral contribution
from the trap region $0\le r\le 0.3R_{\rm F}$, and (2) shows the 
contribution from
the larger trap region 
$0\le r\le 0.6 R_{\rm F}$. 
These results clearly show that tunneling from the low-energy surface states
is the origin of the large low frequency peak. 
The inset in panel (b) shows the profile of 
the computed composite order parameter ${\tilde \Delta}(r)$
used for $\nu=0.6\varepsilon_{\rm F}$.
}
\end{figure}
\begin{figure}
\caption{
\label{fig18}
The solid line shows 
the rf-induced current $I_F(\omega)$ based on the ad-hoc ``broad" composite
order parameter ${\tilde \Delta}(r)$ shown in the inset. This has the same
maximum value ${\tilde \Delta}(0)$ at the center of the trap as the 
correct order parameter shown in the inset of Fig. \ref{fig17}(b).
However, the width is broader. In this calculation, we have also 
used an ad-hoc broad density profile $n_{\rm F}(r)$, with a width of 
the same order as ${\tilde \Delta}(r)$.
The dashed line shows the rf-spectrum plotted in Fig. \ref{fig17}(b), 
based on the self-consistent values of ${\tilde \Delta}(r)$ and $n_{\rm F}(r)$.
}
\end{figure}

\begin{figure}
\caption{
\label{fig19}
Spectrum of ${\tilde I}_{\rm F}(\omega)\equiv 
\omega^2I_{\rm F}(\omega)$ versus the detuning frequency $\omega$, 
for $\nu=0.6\varepsilon_{\rm F}$.
Parameters are the same as in Fig. \ref{fig16}(b).
The arrow shows the value of ${\tilde \Delta}(r=0)$ at 
the center of the trap. The dashed line shows a fit to Eq. (\ref{rf1})
{\it around the high energy region} ($\omega\sim{\tilde \Delta}(0)$).
}
\end{figure}
\begin{figure}
\caption{
\label{fig20}
Molecular dissociation current $I_M(\omega)$ as a function of the
effective detuning frequency $\omega$.
The arrow shows the value of the composite
order parameter ${\tilde \Delta}(0)$ at 
the center of the trap. Parameters are the same as in Fig. \ref{fig19}.
}
\end{figure}
\vskip20mm
\begin{figure}
\caption{
\label{fig21}
Peak energy in the rf-tunneling current. In the crossover
regime ($0.2<\mu/\varepsilon_{\rm F}<1$), the current is peaked at an
energy evaluated using Eq. (\ref{rf1}). The solid line shows ${\tilde \Delta}(0)$ 
in the case of ${\bar U}=0.35\varepsilon_{\rm F}$ (see also Fig. \ref{fig13}).
Experimental data (open circles) are taken from Figs. 1 and 2 of Ref. \cite{Chin}. 
}
\end{figure}
\begin{figure}
\caption{
\label{fig22}
Calculated frequency of the monopole mode 
($L=0$ and $n=1$) in the BCS-BEC crossover region at $T=0$. 
In this figure, $2E_{\rm g}$ describes the threshold of the two-particle
continuum spectrum. The monopole mode is suppressed below
this two-particle continuum.
The collective mode frequency is obtained from the pole of density-density
correction function calculated in the generalized random 
phase approximation\cite{Ohashi-web,Ohashi7}.
}
\end{figure}


\begin{thebibliography}{99}
\bibitem{note4} Since $H_{\rm HFB}^{lm}$ is independent of 
the quantum number $m$, we need only consider $H_{\rm HFB}^{l0}$.
\bibitem{notenum4}
The slight difference between the two is due to the small difference of the chemical 
potential determined by the equation for the total number of atoms, as shown in 
Fig. \ref{fig2}(a).
\bibitem{notePRL} This ``excitation" energy gap is measured relative to 
the value of the chemical
potential $\mu$. In contrast, Fig. 3 of Ref. \cite{Ohashi6} 
gives a plot of $\mu+E_g$.
\bibitem{note0704} This value of the excitation threshold is rather larger than 
$E_g\simeq0.04\varepsilon_{\rm F}$ given in Fig. \ref{fig11}. The difference 
is simply due to the fact that the threshold energy ($\equiv E_{\rm th}$) in the
tunneling current $I_F(\omega)$ is determined by 
$E_{\rm th}=Min[\xi_{nl}^F+E_{jl}^F]=Min[\xi_{nl}^F]+E_g$, under the condition
that energy of the 
$\Psi_a$-state [$\xi_{nl}^F=\hbar\omega_0(2n+l+3/2)-\mu$] is positive. 
In panel (a) of Fig. \ref{fig16}, the chemical potential is
$\mu=27.56\omega_0$. 
Choosing the state with quantum numbers $(n,l)$ such that $2n+l=27$, 
$Min[\xi_{nl}^F]\simeq 0.94\omega_0\simeq 0.03\varepsilon_{\rm F}$.
This gives a tunneling threshold energy 
$E_{\rm th}=E_g+Min[\xi_{nl}^F]\simeq 0.04\varepsilon_{\rm F}+0.03\varepsilon_{\rm F}
=0.07\varepsilon_{\rm F}\simeq 0.08\varepsilon_{\rm F}$ in agreement 
with Fig. \ref{fig16}(a).
\end{thebibliography}
\end{document}